
%
\def\unredoffs{}
\tolerance=1000\hfuzz=2pt
\catcode`\@=11 
\ifx\hyperdef\UNd@FiNeD\def\hyperdef#1#2#3#4{#4}\def\hyperref#1#2#3#4{#4}\def\href#1#2{#2}\fi
\magnification=1200\unredoffs\baselineskip=16pt plus 2pt minus 1pt
\def\Date#1{\vfill\leftline{#1}\tenpoint\supereject%
\footline={\hss\tenrm\hyperdef\hypernoname{page}\folio\folio\hss}}%

{\count255=\time\divide\count255 by 60 \xdef\hourmin{\number\count255}
 \multiply\count255 by-60\advance\count255 by\time
 \xdef\hourmin{\hourmin:\ifnum\count255<10 0\fi\the\count255}
}
\def\date{\number\day.\number\month.\number\year\ at \hourmin}


\def\nolabels{\def\wrlabeL##1{}\def\eqlabeL##1{}\def\reflabeL##1{}}
\def\writelabels{\def\wrlabeL##1{\leavevmode\vadjust{\rlap{\smash%
{\line{{\escapechar=` \hfill\rlap{\sevenrm\hskip.03in\string##1}}}}}}}%
\def\eqlabeL##1{{\escapechar-1\rlap{\sevenrm\hskip.05in\string##1}}}%
\def\reflabeL##1{\noexpand\llap{\noexpand\sevenrm\string\string\string##1}}}
\nolabels

\global\newcount\secno \global\secno=0
\global\newcount\meqno \global\meqno=1
\def\s@csym{}

\def\newsec#1\par{\global\advance\secno by1%
{\toks0{#1}\message{(\the\secno. \the\toks0)}}%
\global\subsecno=0\eqnres@t\let\s@csym\secsym\xdef\secn@m{\the\secno}\noindent
{\bf\hyperdef\hypernoname{section}{\the\secno}{\the\secno.} #1}%
\writetoca{{\string\hyperref{}{section}{\the\secno}{\bf \the\secno\quad}} {\bf #1}}\par%
\nobreak\medskip\nobreak\noindent\ignorespaces}
\def\eqnres@t{\xdef\secsym{\the\secno.}\global\meqno=1\bigbreak\bigskip}
\def\sequentialequations{\def\eqnres@t{\bigbreak}}\xdef\secsym{}

\global\newcount\subsecno \global\subsecno=0
\def\subsec#1\par{\global\advance\subsecno by1%
{\toks0{#1}\message{(\s@csym\the\subsecno. \the\toks0)}}%
\global\subsubsecno=0%
\ifnum\lastpenalty>9000\else\bigbreak\fi
\noindent{\it\hyperdef\hypernoname{subsection}{\secn@m.\the\subsecno}%
{\secn@m.\the\subsecno.} #1}\writetoca{\string\hskip1.45cm
{\string\hyperref{}{subsection}{\secn@m.\the\subsecno}{\secn@m.\the\subsecno.}}
{#1}}\par\nobreak\medskip\nobreak\noindent\ignorespaces}

\def\appendix#1#2{\global\meqno=1\global\subsecno=0\xdef\secsym{\hbox{#1.}}%
\bigbreak\bigskip\noindent{\bf Appendix \hyperdef\hypernoname{appendix}{#1}%
{#1.} #2}{\toks0{(#1. #2)}\message{\the\toks0}}%
\xdef\s@csym{#1.}\xdef\secn@m{#1}%
\writetoca{{\string\hyperref{}{appendix}{#1}{\bf {#1}\quad}} {\bf #2}}%
\par\nobreak\medskip\nobreak}

%
\def\checkm@de#1#2{\ifmmode{\def\f@rst##1{##1}\hyperdef\hypernoname{equation}%
{#1}{#2}}\else\hyperref{}{equation}{#1}{#2}\fi}
\def\eqnn#1{\DefWarn#1\xdef #1{(\noexpand\relax\noexpand\checkm@de%
{\s@csym\the\meqno}{\secsym\the\meqno})}%
\wrlabeL#1\writedef{#1\leftbracket#1}\global\advance\meqno by1}
\def\f@rst#1{\c@t#1a\em@ark}\def\c@t#1#2\em@ark{#1}
\def\eqna#1{\DefWarn#1\wrlabeL{#1$\{\}$}%
\xdef #1##1{(\noexpand\relax\noexpand\checkm@de%
{\s@csym\the\meqno\noexpand\f@rst{##1}1}{\hbox{$\secsym\the\meqno##1$}})}
\writedef{#1\numbersign1\leftbracket#1{\numbersign1}}\global\advance\meqno by1}
\def\eqn#1#2{\DefWarn#1%
\xdef #1{(\noexpand\hyperref{}{equation}{\s@csym\the\meqno}%
{\secsym\the\meqno})}$$#2\eqno(\hyperdef\hypernoname{equation}%
{\s@csym\the\meqno}{\secsym\the\meqno})\eqlabeL#1$$%
\writedef{#1\leftbracket#1}\global\advance\meqno by1}
\def\xeqn{\expandafter\xe@n}\def\xe@n(#1){#1}
\def\xeqna#1{\expandafter\xe@n#1}
\def\eqns#1{(\e@ns #1{\hbox{}})}
\def\e@ns#1{\ifx\UNd@FiNeD#1\message{eqnlabel \string#1 is undefined.}%
\xdef#1{(?.?)}\fi{\let\hyperref=\relax\xdef\next{#1}}%
\ifx\next\em@rk\def\next{}\else%
\ifx\next#1\xeqn#1\else\def\n@xt{#1}\ifx\n@xt\next#1\else\xeqna#1\fi
\fi\let\next=\e@ns\fi\next}

\def\DefWarn#1{\ifx\UNd@FiNeD#1\else
\immediate\write16{*** WARNING: the label \string#1 is already defined ***}\fi}
%
\newskip\footskip\footskip14pt plus 1pt minus 1pt 
\def\footnotefont{\ninepoint}\def\f@t#1{\footnotefont #1\@foot}
\def\f@@t{\baselineskip\footskip\bgroup\footnotefont\aftergroup\@foot\let\next}
\setbox\strutbox=\hbox{\vrule height9.5pt depth4.5pt width0pt}
\global\newcount\ftno \global\ftno=0
\def\foot{\global\advance\ftno by1\def\foot@rg{\hyperref{}{footnote}%
{\the\ftno}{\the\ftno}\xdef\foot@rg{\noexpand\hyperdef\noexpand\hypernoname%
{footnote}{\the\ftno}{\the\ftno}}}\footnote{$^{\foot@rg}$}}
%
%
%
\global\newcount\refno \global\refno=1
\newwrite\rfile
\def\ref{[\hyperref{}{reference}{\the\refno}{\the\refno}]\nref}
\def\nref#1{\DefWarn#1%
\xdef#1{[\noexpand\hyperref{}{reference}{\the\refno}{\the\refno}]}%
\writedef{#1\leftbracket#1}%
\ifnum\refno=1\immediate\openout\rfile=\jobname.refs\fi
\chardef\wfile=\rfile\immediate\write\rfile{\noexpand\item{[\noexpand\hyperdef%
\noexpand\hypernoname{reference}{\the\refno}{\the\refno}]\ }%
\reflabeL{#1\hskip.31in}\pctsign}\global\advance\refno by1\findarg}
\def\findarg#1#{\begingroup\obeylines\newlinechar=`\^^M\pass@rg}
{\obeylines\gdef\pass@rg#1{\writ@line\relax #1^^M\hbox{}^^M}%
\gdef\writ@line#1^^M{\expandafter\toks0\expandafter{\striprel@x #1}%
\edef\next{\the\toks0}\ifx\next\em@rk\let\next=\endgroup\else\ifx\next\empty%
\else\immediate\write\wfile{\the\toks0}\fi\let\next=\writ@line\fi\next\relax}}
\def\striprel@x#1{} \def\em@rk{\hbox{}}
\def\lref{\begingroup\obeylines\lr@f}
\def\lr@f#1#2{\DefWarn#1\gdef#1{\let#1=\UNd@FiNeD\ref#1{#2}}\endgroup\unskip}
\def\semi{;\hfil\break}
\def\addref#1{\immediate\write\rfile{\noexpand\item{}#1}} 
\def\listrefs{\vfill\supereject\immediate\closeout\rfile\writestoppt
\baselineskip=\footskip\centerline{{\bf References}}\bigskip{\parindent=20pt%
\frenchspacing\escapechar=` \input \jobname.refs\vfill\eject}\nonfrenchspacing}
\def\startrefs#1{\immediate\openout\rfile=\jobname.refs\refno=#1}
\def\xref{\expandafter\xr@f}\def\xr@f[#1]{#1}
\def\refs#1{\count255=1[\r@fs #1{\hbox{}}]}
\def\r@fs#1{\ifx\UNd@FiNeD#1\message{reflabel \string#1 is undefined.}%
\nref#1{need to supply reference \string#1.}\fi%
\vphantom{\hphantom{#1}}{\let\hyperref=\relax\xdef\next{#1}}%
\ifx\next\em@rk\def\next{}%
\else\ifx\next#1\ifodd\count255\relax\xref#1\count255=0\fi%
\else#1\count255=1\fi\let\next=\r@fs\fi\next}
%

%
\newwrite\ffile\global\newcount\figno \global\figno=1
\def\fig{fig.~\hyperref{}{figure}{\the\figno}{\the\figno}\nfig}
\def\nfig#1{\DefWarn#1%
\xdef#1{fig.~\noexpand\hyperref{}{figure}{\the\figno}{\the\figno}}%
\writedef{#1\leftbracket fig.\noexpand~\xfig#1}%
\ifnum\figno=1\immediate\openout\ffile=\jobname.figs\fi\chardef\wfile=\ffile%
{\let\hyperref=\relax
\immediate\write\ffile{\noexpand\medskip\noexpand\item{Fig.\ %
\noexpand\hyperdef\noexpand\hypernoname{figure}{\the\figno}{\the\figno}. }
\reflabeL{#1\hskip.55in}\pctsign}}\global\advance\figno by1\findarg}
\def\xfig{\expandafter\xf@g}\def\xf@g fig.\penalty\@M\ {}
\def\figs#1{figs.~\f@gs #1{\hbox{}}}
\def\f@gs#1{{\let\hyperref=\relax\xdef\next{#1}}\ifx\next\em@rk\def\next{}\else
\ifx\next#1\xfig #1\else#1\fi\let\next=\f@gs\fi\next}
%
\def\figin{\epsfcheck\figin}\def\figins{\epsfcheck\figins}
\def\epsfcheck{\ifx\epsfbox\UnDeFiNeD
\message{(NO epsf.tex, FIGURES WILL BE IGNORED)}
\gdef\figin##1{\vskip2in}\gdef\figins##1{\hskip.5in}
\else\message{(FIGURES WILL BE INCLUDED)}%
\gdef\figin##1{##1}\gdef\figins##1{##1}\fi}
\def\DefWarn#1{}
\def\figinsert{\goodbreak\topinsert}
\def\ifig#1#2#3{\DefWarn#1\xdef#1{fig.~\the\figno}
\writedef{#1\leftbracket fig.\noexpand~\the\figno}%
\figinsert\figin{\centerline{#3}}
\smallskip
\leftskip=20pt \rightskip=20pt
\baselineskip12pt\noindent
{{\bf Fig.~\the\figno}\ \ninepoint #2}
\medskip
\global\advance\figno by1\par\endinsert}
\newwrite\lfile
{\escapechar-1\xdef\pctsign{\string\%}\xdef\leftbracket{\string\{}
\xdef\rightbracket{\string\}}\xdef\numbersign{\string\#}}
\def\writedefs{\immediate\openout\lfile=label.defs \def\writedef##1{%
{\let\hyperref=\relax\let\hyperdef=\relax\let\hypernoname=\relax
 \immediate\write\lfile{\string\def\string##1\rightbracket}}}}%
\def\writestop{\def\writestoppt{\immediate\write\lfile{\string\pageno
 \the\pageno\string\startrefs\leftbracket\the\refno\rightbracket
 \string\def\string\secsym\leftbracket\secsym\rightbracket
 \string\secno\the\secno\string\meqno\the\meqno}\immediate\closeout\lfile}}
\def\writestoppt{}\def\writedef#1{}

\def\seclab#1{\DefWarn#1%
\xdef #1{\noexpand\hyperref{}{section}{\the\secno}{\the\secno}}%
\writedef{#1\leftbracket#1}\wrlabeL{#1=#1}}
\def\subseclab#1{\DefWarn#1%
\xdef #1{\noexpand\hyperref{}{subsection}{\the\secno.\the\subsecno}%
{\the\secno.\the\subsecno}}\writedef{#1\leftbracket#1}\wrlabeL{#1=#1}}
\def\applab#1{\DefWarn#1%
\xdef #1{\noexpand\hyperref{}{appendix}{\secn@m}{\secn@m}}%
\writedef{#1\leftbracket#1}\wrlabeL{#1=#1}}
\newwrite\tfile \def\writetoca#1{}
\def\leaderfill{\leaders\hbox to 1em{\hss.\hss}\hfill}
\def\writetoc{\immediate\openout\tfile=\jobname.toc
   \def\writetoca##1{{\edef\next{\write\tfile{\noindent ##1
   \string\leaderfill{
   \string\hyperref{}{page}{\noexpand\number\pageno}%
   {\noexpand\number\pageno}} \par}}\next}}
}
\newread\ch@ckfile
\def\listtoc{\immediate\closeout\tfile\immediate\openin\ch@ckfile=\jobname.toc
\ifeof\ch@ckfile\message{no file \jobname.toc, no table of contents this pass}%
\else\closein\ch@ckfile\centerline{\bf Contents}\nobreak\medskip%
{\baselineskip=16pt\footnotefont\parskip=0pt\catcode`\@=11\input\jobname.toc
\catcode`\@=12\bigbreak\bigskip}\fi}
\catcode`\@=12 
\def\tenpoint{\def\rm{\fam0\tenrm}
\textfont0=\tenrm \scriptfont0=\sevenrm \scriptscriptfont0=\fiverm
\textfont1=\teni  \scriptfont1=\seveni  \scriptscriptfont1=\fivei
\textfont2=\tensy \scriptfont2=\sevensy \scriptscriptfont2=\fivesy
\textfont\itfam=\tenit \def\it{\fam\itfam\tenit}\def\footnotefont{\ninepoint}%
\textfont\bffam=\tenbf \def\bf{\fam\bffam\tenbf}\def\sl{\fam\slfam\tensl}\rm}
\font\ninerm=cmr9 \font\sixrm=cmr6 \font\ninei=cmmi9 \font\sixi=cmmi6
\font\ninesy=cmsy9 \font\sixsy=cmsy6 \font\ninebf=cmbx9
\font\nineit=cmti9 \font\ninesl=cmsl9 \skewchar\ninei='177
\skewchar\sixi='177 \skewchar\ninesy='60 \skewchar\sixsy='60
\def\ninepoint{\def\rm{\fam0\ninerm}
\textfont0=\ninerm \scriptfont0=\sixrm \scriptscriptfont0=\fiverm
\textfont1=\ninei \scriptfont1=\sixi \scriptscriptfont1=\fivei
\textfont2=\ninesy \scriptfont2=\sixsy \scriptscriptfont2=\fivesy
\textfont\itfam=\ninei \def\it{\fam\itfam\nineit}\def\sl{\fam\slfam\ninesl}%
\textfont\bffam=\ninebf \def\bf{\fam\bffam\ninebf}\rm}
%
\hyphenation{anom-aly anom-alies coun-ter-term coun-ter-terms}

\global\newcount\subsubsecno \global\subsubsecno=0
\def\subsubsec#1\par{\global\advance\subsubsecno by1%
{\toks0{#1}\message{(\the\secno\the\subsecno\the\subsubsecno. \the\toks0)}}%
\ifnum\lastpenalty>9000\else\bigbreak\fi
\noindent{\it\hyperdef\hypernoname{subsubsection}{\the\secno.\the\subsecno\the\subsubsecno}%
{\the\secno.\the\subsecno.\the\subsubsecno.} #1}
\par\nobreak\medskip\nobreak\noindent\ignorespaces}

\def\DefWarn#1{}
\def\tikzcaption#1#2{\DefWarn#1\xdef#1{Fig.~\the\figno}
\writedef{#1\leftbracket Fig.\noexpand~\the\figno}%
{
\smallskip
\leftskip=20pt \rightskip=20pt \baselineskip12pt\noindent
{{\bf Fig.~\the\figno}\ \ninepoint #2}
\bigskip
\global\advance\figno by1 \par}}

\def\ntoalpha#1{%
\ifcase#1%
@%
\or A\or B\or C\or D\or E\or F\or G\or H\or I
\fi
}

\global\newcount\appno \global\appno=1
\def\applab#1{\xdef #1{\ntoalpha\appno}\writedef{#1\leftbracket#1}\wrlabeL{#1=#1}
\global\advance\appno by1}

\def\preprint#1 #2\par{\rightline{\vbox{\baselineskip12pt\hbox{#1}\hbox{#2}}}\vskip2cm}
%
\def\title#1\par{\centerline{\bf #1}\nopagenumbers\pageno=0}
\def\author#1\par{\bigskip\bigskip\centerline{#1}}

\newcount\addressno

\def\email#1#2{\unskip$^#1$\footnote{\null}{\kern-\parindent \llap{$^#1$\hskip1pt}email: #2}}

\def\startcenter{%
  \par
  \begingroup
  \leftskip=0pt plus 1fil
  \rightskip=\leftskip
  \parindent=0pt
  \parfillskip=0pt
}
\def\stopcenter{\endgroup}

\def\address{\bigskip%
  \ifnum\the\addressno=0\else\stopcenter\endgroup\fi
  \advance\addressno by 1%
  \begingroup
  \startcenter
  \it
  \obeylines
  \addressAux
}
\def\addressAux#1{#1}

\def\abstract{\stopcenter\endgroup\bigskip\bigskip\noindent}

\def\Dsl{\,\raise.15ex\hbox{/}\mkern-13.5mu D} 
\def\dsl{\raise.15ex\hbox{/}\kern-.57em\partial}
 
\def\boxeqn#1{\vcenter{\vbox{\hrule\hbox{\vrule\kern3pt\vbox{\kern3pt
	\hbox{${\displaystyle #1}$}\kern3pt}\kern3pt\vrule}\hrule}}}


\def\ap{{\alpha^{\prime}}}
\def\halfap#1{\Big({\ap\over 2}\Big)^{\mkern-4mu #1}}
\def\a{\alpha}
\def\b{{\beta}}
\def\g{{\gamma}}
\def\d{{\delta}}
\def\e{{\epsilon}}
\def\l{\lambda}
\def\k{{\kappa}}
\def\s{{\sigma}}
\def\t{{\theta}}

\def\lb{{\overline\lambda}}
\def\llb{(\l\lb)}
\def\wb{{\overline w}}
\def\half{{1\over 2}}
\def\p{{\partial}}
\def\pb{{\overline\partial}}

\def\bar{\overline}
\def\({\left(}
\def\){\right)}
\def\cF{{\cal F}}

\def\eikx{{\bigl\langle \prod_{j=1}^4 {\rm e}^{i k^j\cdot x^j}\bigr\rangle}}
\def\ImOmega{\mathop{\rm Im}\Omega}

\def\Im{\mathop{{\rm Im}}} 


\def\qed{\hbox{\hskip 3pt
\vbox{\hrule\hbox to 7pt{\vrule height 7pt\hfill\vrule}
\hrule}}\hskip3pt}

\overfullrule=0pt\relax

\frenchspacing

\newread\instream \openin\instream= label.defs
\ifeof\instream \message{No labels in advance yet. Wait till next pass.}
\else \closein\instream \input label.defs
\fi
\writedefs

\def\arXiv:#1].{\hepthStrip#1 \nil}
\def\hepthStrip#1 #2\nil{\href{http://arxiv.org/abs/#1}{arXiv:#1 #2\unskip}].}

\input amssym.tex 

\def\la#1{{\lambda^{\a_{#1}}}}
\def\DeltaPi#1,#2,#3,#4,#5{\Delta^{#5}(z_{#1},z_{#2};z_{#3};z_{#4})}
\def\DeltaPibar#1,#2,#3,#4,#5{\bar\Delta^{#5}(\bar z_{#1},\bar z_{#2};\bar z_{#3};\bar z_{#4})}
\def\eikx#1{\bigl\langle\prod_{j=1}^{#1} e^{i k^j\cdot x^j}\bigr\rangle}
\font\titlefont=cmr10 scaled\magstep3
\def\title#1\par{\centerline{\titlefont #1}\nopagenumbers\pageno=0}

\preprint DAMTP--2013--50

\title The closed-string 3-loop amplitude and S-duality

\author Humberto Gomez\email{\dagger}{humgomzu@ift.unesp.br} and
        Carlos R. Mafra\email{\ddagger}{c.r.mafra@damtp.cam.ac.uk}

\address
$^\dagger$Perimeter Institute for Theoretical Physics, Waterloo, Ontario N2L 2Y5, Canada
$^\dagger$Instituto de F\'\i sica Te\'orica UNESP --- Universidade Estadual Paulista
Caixa Postal 70532--2 01156--970 S\~ao Paulo, SP, Brazil

\address
$^\ddagger$DAMTP, University of Cambridge
Wilberforce Road, Cambridge, CB3 0WA, UK

\abstract
The low-energy limit of the four-point 3-loop amplitude (including its overall coefficient)
is computed in both type IIA and IIB superstring theories using the pure spinor formalism.
The result is shown to agree with the prediction of the coefficient for the type IIB $D^6 R^4$ interaction made by
Green and Vanhove based on S-duality considerations.

\Date {August 2013}


\lref\renann{
  R.~Lipinski Jusinskas,
  ``Nilpotency of the b ghost in the non-minimal pure spinor formalism,''
JHEP {\bf 1305}, 048 (2013).
[arXiv:1303.3966 [hep-th]].
}

\lref\sakai{
  N.~Sakai and Y.~Tanii,
  ``One Loop Amplitudes And Effective Action In Superstring Theories,''
Nucl.\ Phys.\ B {\bf 287}, 457 (1987)..
}

\lref\SS{
  O.~Schlotterer and S.~Stieberger,
  ``Motivic Multiple Zeta Values and Superstring Amplitudes,''
[arXiv:1205.1516 [hep-th]].
}

\lref\anomalia{
	N.~Berkovits and C.R.~Mafra,
	``Some superstring amplitude computations with the non-minimal pure spinor
	formalism,''
	JHEP {\bf 0611}, 079 (2006)
	[arXiv:hep-th/0607187].
}

\lref\NathanTheorem{
	N.~Berkovits,
	``New higher-derivative R**4 theorems,''
	Phys.\ Rev.\ Lett.\  {\bf 98}, 211601 (2007).
	[arXiv:hep-th/0609006].
}

\lref\GreenDhoker{
	E.~D'Hoker and M.~B.~Green,
	``Zhang-Kawazumi Invariants and Superstring Amplitudes,''
	[arXiv:1308.4597 [hep-th]].
}

\lref\nptSYM{
	C.R.~Mafra, O.~Schlotterer, S.~Stieberger and D.~Tsimpis,
	``A recursive method for SYM n-point tree amplitudes,''
	Phys.\ Rev.\ D {\bf 83}, 126012 (2011).
	[arXiv:1012.3981 [hep-th]].
}
\lref\oneloopbb{
	C.R.~Mafra and O.~Schlotterer,
	``The Structure of n-Point One-Loop Open Superstring Amplitudes,''
	[arXiv:1203.6215 [hep-th]].
}
\lref\psf{
	N.~Berkovits,
	``Super-Poincare covariant quantization of the superstring,''
	JHEP {\bf 0004}, 018 (2000)
	[arXiv:hep-th/0001035].
}
\lref\multiloop{
	N.~Berkovits,
	``Multiloop amplitudes and vanishing theorems using the pure spinor
	formalism for the superstring,''
	JHEP {\bf 0409}, 047 (2004)
	[arXiv:hep-th/0406055].
}
\lref\twoloop{
	N.~Berkovits,
	``Super-Poincare covariant two-loop superstring amplitudes,''
	JHEP {\bf 0601}, 005 (2006)
	[arXiv:hep-th/0503197].
}
\lref\twolooptwo{
  N.~Berkovits and C.R.~Mafra,
  ``Equivalence of two-loop superstring amplitudes in the pure spinor and  RNS
  formalisms,''
  Phys.\ Rev.\ Lett.\  {\bf 96}, 011602 (2006)
  [arXiv:hep-th/0509234].
}
\lref\nekrasov{
  N.~Berkovits and N.~Nekrasov,
  ``Multiloop superstring amplitudes from non-minimal pure spinor formalism,''
  JHEP {\bf 0612}, 029 (2006)
  [arXiv:hep-th/0609012].
}
\lref\PSSuperspace{
  N.~Berkovits,
  ``Explaining pure spinor superspace,''
  [arXiv:hep-th/0612021].
}
\lref\NMPS{
  N.~Berkovits,
  ``Pure spinor formalism as an N = 2 topological string,''
  JHEP {\bf 0510}, 089 (2005)
  [arXiv:hep-th/0509120].
}
\lref\humberto{
  H.~Gomez,
  ``One-loop Superstring Amplitude From Integrals on Pure Spinors Space,''
  JHEP {\bf 0912}, 034 (2009)
  [arXiv:0910.3405 [hep-th]].
}
\lref\dhokerbosonic{
  E.~D'Hoker and D.~H.~Phong,
  ``Multiloop Amplitudes for the Bosonic Polyakov String,''
Nucl.\ Phys.\ B {\bf 269}, 205 (1986)..
}
\lref\dhokerVI{
  E.~D'Hoker and D.H.~Phong,
  ``Two-Loop Superstrings VI: Non-Renormalization Theorems and the 4-Point
  Function,''
  Nucl.\ Phys.\  B {\bf 715}, 3 (2005)
  [arXiv:hep-th/0501197].
}
\lref\dhokerS{
  E.~D'Hoker, M.~Gutperle and D.H.~Phong,
  ``Two-loop superstrings and S-duality,''
  Nucl.\ Phys.\  B {\bf 722}, 81 (2005)
  [arXiv:hep-th/0503180].
}

\lref\FORM{
	J.A.M.~Vermaseren,
	``New features of FORM,''
	[arXiv:math-ph/0010025].
}

\lref\dhokerReview{
 E.~D'Hoker and D.H.~Phong,
 ``The Geometry of String Perturbation Theory,''
  Rev.\ Mod.\ Phys.\  {\bf 60}, 917 (1988).
}

\lref\bOda{
  I.~Oda and M.~Tonin,
  ``Y-formalism and $b$ ghost in the Non-minimal Pure Spinor Formalism of
  Superstrings,''
  Nucl.\ Phys.\  B {\bf 779}, 63 (2007)
  [arXiv:0704.1219 [hep-th]].
}

\lref\fiveone{
  C.R.~Mafra and C.~Stahn,
  ``The One-loop Open Superstring Massless Five-point Amplitude with the
  Non-Minimal Pure Spinor Formalism,''
  JHEP {\bf 0903}, 126 (2009)
  [arXiv:0902.1539 [hep-th]].
}
\lref\verlinde{
  E.P.~Verlinde and H.L.~Verlinde,
  ``Chiral bosonization, determinants and the string partition function,''
  Nucl.\ Phys.\  B {\bf 288}, 357 (1987).
}
\lref\GreenOneLoopcoeff{
  M.B.~Green, J.G.~Russo and P.~Vanhove,
  ``Low energy expansion of the four-particle genus-one amplitude in type II
  superstring theory,''
  JHEP {\bf 0802}, 020 (2008)
  [arXiv:0801.0322 [hep-th]].
}
\lref\harris{
Griffiths and Harris, 
``Principles of Algebraic Geometry'', [Wiley Classics Library Edition
Published 1994]
}

\lref\coefftwo{
	H.~Gomez, C.R.~Mafra,
	``The Overall Coefficient of the Two-loop Superstring Amplitude Using Pure Spinors,''
	JHEP {\bf 1005}, 017 (2010).
	[arXiv:1003.0678 [hep-th]].
}
\lref\siegel{
	W.~Siegel,
	``Classical Superstring Mechanics,''
	Nucl.\ Phys.\  {\bf B263}, 93 (1986).
}
\lref\refSYM{
	E.Witten,
        ``Twistor-Like Transform In Ten-Dimensions''
        Nucl.Phys. B {\bf 266}, 245~(1986)
}

\lref\HoogeveenHK{
	J.~Hoogeveen and K.~Skenderis,
	``Decoupling of unphysical states in the minimal pure spinor formalism I,''
	JHEP {\bf 1001}, 041 (2010).
	[arXiv:0906.3368 [hep-th]].
}
\lref\aldoPF{
	Y.~Aisaka, E.A.~Arroyo, N.~Berkovits and N.~Nekrasov,
	``Pure Spinor Partition Function and the Massive Superstring Spectrum,''
	JHEP {\bf 0808}, 050 (2008)
	[arXiv:0806.0584 [hep-th]].
}
\lref\character{
	N.~Berkovits and N.~Nekrasov,
	``The Character of pure spinors,''
	Lett.\ Math.\ Phys.\  {\bf 74}, 75 (2005).
	[arXiv:hep-th/0503075].
}
\lref\treebbI{
	C.R.~Mafra, O.~Schlotterer and S.~Stieberger,
	``Complete N-Point Superstring Disk Amplitude I. Pure Spinor Computation,''
	Nucl.\ Phys.\ B {\bf 873}, 419 (2013).
	[arXiv:1106.2645 [hep-th]].
 \semi
  	C.R.~Mafra, O.~Schlotterer and S.~Stieberger,
  	``Complete N-Point Superstring Disk Amplitude II. Amplitude and Hypergeometric Function Structure,''
	Nucl.\ Phys.\ B {\bf 873}, 461 (2013).
	[arXiv:1106.2646 [hep-th]].
}

\lref\Greenthreeloop{
	M.B.~Green and P.~Vanhove,
	``Duality and higher derivative terms in M theory,''
	JHEP {\bf 0601}, 093 (2006).
	[arXiv:hep-th/0510027].
}
\lref\SiegelVol{
	C.L. Siegel, ``Symplectic Geometry'', Am. J. Math. 65 (1943) 1-86;
}
\lref\GGRq{
	M.B.~Green and M.~Gutperle,
	``Effects of D instantons,''
	Nucl.\ Phys.\ B {\bf 498}, 195 (1997).
	[hep-th/9701093].
\semi
	M.B.~Green, M.~Gutperle and P.~Vanhove,
	``One loop in eleven-dimensions,''
	Phys.\ Lett.\ B {\bf 409}, 177 (1997).
	[hep-th/9706175].
}
\lref\GreenKVan{
	M.B.~Green, H.-h.~Kwon and P.~Vanhove,
	``Two loops in eleven-dimensions,''
	Phys.\ Rev.\ D {\bf 61}, 104010 (2000).
	[hep-th/9910055].
}
\lref\stahnCorr{
	C.~Stahn,
	``Fermionic superstring loop amplitudes in the pure spinor formalism,''
	JHEP {\bf 0705}, 034 (2007).
	[arXiv:0704.0015 [hep-th]].
}
\lref\PSS{
	C.R.~Mafra,
	``PSS: A FORM Program to Evaluate Pure Spinor Superspace Expressions,''
	[arXiv:1007.4999 [hep-th]].
}

\lref\FiveSduality{
	M.B.~Green, C.R.~Mafra and O.~Schlotterer,
	``Multiparticle one-loop amplitudes and S-duality in closed superstring theory,''
	[arXiv:1307.3534 [hep-th]].
}
\lref\vafa{
  M.~Bershadsky, S.~Cecotti, H.~Ooguri and C.~Vafa,
  ``Kodaira-Spencer theory of gravity and exact results for quantum string amplitudes,''
Commun.\ Math.\ Phys.\  {\bf 165}, 311 (1994).
[hep-th/9309140].
}

\lref\yuri{
  Y.~Aisaka and N.~Berkovits,
  ``Pure Spinor Vertex Operators in Siegel Gauge and Loop Amplitude Regularization,''
JHEP {\bf 0907}, 062 (2009).
[arXiv:0903.3443 [hep-th]].
}

\lref\WittenCIA{
  E.~Witten,
  ``More On Superstring Perturbation Theory,''
[arXiv:1304.2832 [hep-th]].
}
\lref\BerkovitsAW{
  N.~Berkovits, M.B.~Green, J.G.~Russo and P.~Vanhove,
  ``Non-renormalization conditions for four-gluon scattering in supersymmetric string and field theory,''
JHEP {\bf 0911}, 063 (2009).
[arXiv:0908.1923 [hep-th]].
}

\listtoc
\writetoc
\filbreak


\newsec{Introduction}

Up to this day superstring amplitudes in ten-dimensional Minkowski space have never been computed at genus higher than two.
In this paper the low-energy limit of the genus three amplitude for four massless states in closed superstring
theory is computed (including its overall coefficient) using the pure spinor formalism \refs{\psf,\NMPS}.

After the relatively straightforward pure spinor derivation of the two-loop amplitude\foot{For the RNS derivation see the earlier works of D'Hoker and Phong, e.g. \dhokerVI.}
in \refs{\twoloop,\twolooptwo},
the natural question was how well the formalism would behave at higher genus. It is well-known by now that, in order to compute general amplitudes at genus
higher than two the original
BRST-invariant regulator of Berkovits \NMPS\ needs to be replaced by a more complicated scheme proposed by Berkovits
and Nekrasov in \nekrasov. Nevertheless, for four massless states at genus three one can still use the original regulator for the
terms considered in this paper since they are F-terms and these were shown in \NathanTheorem\ to be unaffected by the divergences
which require the new regulator.

In addition to the regulator, there is one more point to consider though. As recently emphasized by Witten \WittenCIA, to compute multiloop
scattering amplitudes it is
not sufficient to represent the external states by BRST-invariant vertex operators of definite conformal weight. The unintegrated vertex may
have at most a simple pole singularity with the $b$-ghost while the integrated vertex must have no singularities at all. Unfortunately
this is not the case for the massless pure spinor vertex operators of \refs{\psf,\NMPS} and one would probably need
to use the vertices constructed in \yuri. These vertices depend on the non-minimal variables and therefore require the concomitant use of
the Berkovits--Nekrasov regulator.

Luckily, we will show that the low-energy limit (of order $D^6R^4$) of the genus-three amplitude is not affected by these
considerations because only the zero modes of the $b$-ghost enter in the derivation. Any subtlety is deferred to terms of order
$D^8 R^4$ and higher.

With that in mind, one can proceed with the three-loop computation using the formalism as described in \NMPS. And
ever since the normalizations for the pure spinor measures were determined in \humberto\ and
systematically used in \coefftwo, keeping track of the overall normalization does not pose additional difficulties.
In doing so, the precise normalization of the amplitude at order
$D^6R^4$ is shown to agree with the value predicted
in 2005 by Green and Vanhove using S-duality arguments \Greenthreeloop.

\newsec Definitions and conventions

The non-minimal pure spinor formalism action for the left-moving sector reads \NMPS\
\eqn\action{
S = {1\over 2\pi \ap}\int_{\Sigma_g} d^2z \( \p x^m \pb x_m + \ap p_\a \pb \t^\a
- \ap w_\a \pb\l^\a
- \ap \wb^\a \pb\lb_\a + \ap s^\a\pb r_\a \),
}
where $\l^\a$ and $\lb_\b$ are bosonic pure spinors and $r_\a$ is a constrained fermionic variable,
\eqn\PSconstraints{
(\l\g^m\l) = 0, \quad (\lb\g^m\lb) = 0, \quad (\lb\g^m r) = 0.
}
The fields in \action\ have the following space-time dimensions \humberto
\eqn\dim{
[\ap] = 2,\; [x^m] = 1, \quad
[\t^\a, \l^\a, \wb^\a, s^\a] = 1/2, \quad [p_\a, w_\a, \lb_\a, r_\a] = -1/2.
}
The genus-$g$ OPEs for the matter variables following from \action\ are \verlinde
\eqn\OPEs{
x^m(z,\bar z)\,x_n(w,\bar w) \sim \d^m_n G(z,w), \qquad
p_\a(z)\, \t^\b(w) \sim \d^\b_\a\eta(z,w),
}
where the Green's function $G(z,w)$ is written in terms of the prime form $E(z,w)$ and the global holomorphic 1-forms  $w_I(z)$ as \dhokerReview
\eqn\primeform{
G(z_i, z_j) = -{\ap\over2}\ln\big| E(z_i,z_j)\big|^2
+\ap\pi\, (\Im\!\int_{z_i}^{z_j}\!\!\! w_I)\,(\Im\Omega)^{-1}_{IJ}\,(\Im\!\int_{z_i}^{z_j}\!\!\!w_J),
}
and satisfies
${2\over \ap}\p_{z_i} \pb_{z_j}G(z_i,z_j) = 2\pi\d^{(2)}(z_i-z_j) - \pi\Omega(z_i,z_j),$
where
\eqn\PiOPE{
\Omega(z_i,z_j) \equiv \Omega_{ij}\equiv \sum_{I,J=1}^3 w_I(z_i)(\Im\Omega)_{IJ}^{-1}\bar w_J(\bar z_j)\,,
}
and $\Omega_{IJ}$ is the period matrix which will be defined below. Furthermore,
\eqn\etaij{
\eta(z_i,z_j)=\eta_{ij}\equiv -{2\over \ap}{\p\over \p z_i}G(z_i,z_j).
}
The Green--Schwarz constraint $d_\a(z)$ and the supersymmetric momentum $\Pi^m(z)$ are
\eqn\dalpha{
d_\a = p_\a - {1\over \ap}(\g^m\t)_\a \p x_m - {1\over 4\ap}(\g^m \t)_\a(\t \g_m \p\t), \qquad
\Pi^m = \p x^m + \half (\t\g^m \p\t)
}
and satisfy the following OPEs \siegel
\eqn\opedp{\eqalign{
 d_\a (z)d_\b(w) &\sim - {2\over \ap}{\g^m_{\a\b}\Pi_m \over z-w},\cr
d_\a(z)\Pi^m(w) &\sim {\g^{m}_{\a\b}\p\t^\b \over z-w},
}\quad
\eqalign{
d_\a(z)f(x(w),\t(w)) &\sim {D_{\a}f \over z-w},\cr
\Pi^m(z)f(x(w),\t(w)) &\sim -{\ap\over 2}{k^m f \over z-w}
}}
where
$D_\a = {\p\over \p\t^\a} + \half (\g^m\t)_\a k_m$
is the supersymmetric derivative and $f(x,\t)$ represents a generic superfield. The $b$-ghost is given by \NMPS\ (see also \refs{\bOda,\renann})
\eqnn\bghost
$$\eqalignno{
&\qquad b  = s^\a \p\lb_\a +
{1\over 4(\l\lb)}\bigl[ 2\Pi^{m}(\lb \g_{m}d) -
N_{mn}(\lb \g^{mn}\p\t) - J_{\lambda}(\lb \p\t) - (\lb \p^2 \t) \bigr] &\bghost\cr
&{} + {(\lb\g^{mnp}r)\over 192 (\l\lb)^2}\Bigl[ {\ap\over 2}(d\g_{mnp}d) + 24N_{mn}\Pi_{p}\Bigr]
- {\ap\over 2}{(r\g_{mnp}r) \over 16(\l\lb)^3 }\Bigl[
(\lb\g^{m}d)N^{np}
- {(\lb\g^{pqr}r)N^{mn}N_{qr} \over 8(\l\lb)}\Bigr],
}$$
and satisfies $\{Q,b(z)\} = T(z)$
where the BRST charge $Q$ and the energy-momentum tensor $T(z)$ are
$$
Q=\oint (\l^\a d_\a + \wb^{\a}r_{\a}),\qquad
T(z)=-{1\over \ap}\p x^m \p x_m -
p_\a \p\t^\a + w_{\a}\p\l^{\a} +
\wb^\a \p\lb_{\a} -s^\a\p r_{\a}.
$$
From \dim\ it follows that $[Q]=[b]=[T] = 0$.

The massless vertex operators are given by $V(z,\bar z) = \kappa V(\theta)\otimes {\tilde V}(\bar \theta)\,e^{ik\cdot x}$ and
$U(z,\bar z) = \kappa U(\t)\otimes {\tilde U}(\bar \t)\,e^{ik\cdot x}$, where
\eqn\vertices{
V(z) = \l^\a A_\a, \qquad U(z) = \p\t^\a A_\a + A_m \Pi^m + {\ap\over 2}d_\a W^\a + {\ap \over 4}N_{mn}{\cal F}^{mn}
}
and $A_\a,A^m,W^\a,{\cal F}^{mn}$ are the ${\cal N} = 1$ super-Yang--Mills superfields in ten dimensions satisfying \refSYM
\eqnn\SYM
$$\displaylines{
\hfill D_\a A_\b + D_\b A_\a = \g^m_{\a\b} A_m, \qquad D_\a A_m = (\g_m W)_\a + k_m A_\a  \hfill\phantom{(1.1)}\cr
\hfill D_\a{\cal F}_{mn} = 2k_{[m} (\g_{n]} W)_\a, \qquad  D_\a W^{\b} = {1\over 4}(\g^{mn})_\a{}^\b{\cal F}_{mn}.  \hfill\SYM\cr
}$$
The space-time dimensions of the superfields and the vertex operators are
\eqn\Susydim{
[A_\a] = 1/2, \quad [A_m] = 0, \quad [W^\a] = -1/2, \quad [{\cal F}_{mn}] = -1, \qquad [V(z)] = [U(z)] = 1.
}

\subsec Integration on pure spinor space

The zero-mode measures for the non-minimal pure spinor variables in a genus-$g$ surface
have space-time dimension zero and
are given by \refs{\NMPS,\humberto,\coefftwo}
\eqn\measures{\eqalign{
&[d\l]\, T_{\a_1 \ldots\a_5} = c_{\l}\, \e_{\a_1 \ldots\a_{16}} d\l^{\a_6}\kern-4pt \ldots d\l^{\a_{16}} \cr
&[d\lb]\, {\bar T}^{\a_1 \ldots\a_5} \kern-2pt = c_{\lb}\, \e^{\a_1 \ldots\a_{16}} d\lb_{\a_6} \ldots d\lb_{\a_{16}} \cr
&[dr] = c_r\, {\bar T}^{\a_1 \ldots\a_5} \e_{\a_1 \ldots\a_{16} } \p_{r}^{\a_6} \ldots \p_{r}^{\a_{16}}\cr
&[d\t] = c_\t\, d^{16}\t
}\qquad\eqalign{
&[dw] = c_{w}\, T_{\a_1 \ldots\a_5} \e^{\a_1 \ldots\a_{16}} dw_{\a_6} \ldots dw_{\a_{16}}\cr
&[d\wb]\, T_{\a_1 \ldots\a_5} = c_{\wb}\, \e_{\a_1 \ldots\a_{16}} d\wb^{\a_6}\! \ldots d\wb^{\a_{16}}\cr
&[ds^I] = c_s\, T_{\a_1 \ldots\a_5} \epsilon^{\a_1\ldots \a_{16}} \p^{s^I}_{\a_6}\ldots \p^{s^I}_{\a_{16}}\cr
&[dd^I] = c_d\, d^{16}d^I.
}}
The normalizations are
\eqn\normalizations{\eqalign{
c_{\l} &= \halfap{-2}\mkern-4mu {1 \over 11!} \Big({A_g \over 4\pi^2}\Big)^{\!\! 11/2}\cr
c_{\lb} &= \halfap{2} {2^6 \over 11!} \Big({A_g \over 4\pi^2}\Big)^{\mkern-6mu 11/2}\cr
c_r &= \halfap{-2}\mkern-10mu {R \over 11!5!}\Big({2\pi \over A_g}\Big)^{\mkern-6mu 11/2}\cr
c_{\t} &= \halfap{4}\mkern-4mu\Big({2\pi\over A_g}\Big)^{\mkern-6mu 16/2}\cr
}\qquad\eqalign{
c_{w} &= \halfap{2} {(2\pi)^{-11}\over 11!\,5!}\, Z_g^{-11/g}\cr
c_{\wb} &= \halfap{-2} \mkern-16mu {(\l\lb)^3\over 11!\,(2\pi)^{11}} Z_g^{-11/g}\cr
c_s &= \halfap{2} {(2\pi)^{11/2} R^{-1} \over 2^6 11!\,5!\, (\l\lb)^3} Z_g^{11/g}\cr
c_{d} &= \halfap{-4} \!\! (2\pi)^{16/2}\, Z_g^{16/g}\,. \cr
}}
where $A_g = \int d^2z \sqrt{g}$ is the area of the genus-$g$ Riemann surface and
\eqn\Zg{
 \quad Z_g = {1\over \sqrt{\det(2\ImOmega)}}, \qquad g \ge 1.
}
The tensors $T_{\a_1\ldots \a_5}$ and ${\bar T}^{\a_1\ldots \a_5}$ in \measures,
\eqnn\Ttensors
$$\eqalignno{
T_{\a_1\a_2\a_3\a_4\a_5} &=
(\l \g^m)_{\a_1}(\l \g^n)_{\a_2}(\l \g^p)_{\a_3} (\g_{mnp})_{\a_4\a_5} &\Ttensors\cr
{\bar T}^{\a_1\a_2\a_3\a_4\a_5} &= (\lb \g^m)^{\a_1}(\lb \g^n)^{\a_2}(\lb \g^p)^{\a_3} (\g_{mnp})^{\a_4\a_5}
}$$
are totally antisymmetric due to the pure spinor constraint \PSconstraints\ and satisfy
$T\cdot {\bar T} = 5!\, 2^6 (\l\lb)^3$.
As explained in \coefftwo, setting
$R^2= {\sqrt{2}\over 2^{16}\pi}$ fixes the normalization of pure spinor tree-level amplitudes
to be same as in the RNS computations of \dhokerS.

Using the above measures and the results of \humberto\ one can show that the integration over
an arbitrary number of pure spinors $\l^\a$ and $\lb_\b$ is given by
\eqn\dladlb{
\int[d\l][d\lb]e^{-(\l\lb)}(\l\lb)^m \l^{\a_1}\cdots \l^{\a_n}\lb_{\b_1}\cdots \lb_{\b_n} =
\Big({A_g\over 2\pi}\Big)^{\mkern-6mu 11}{12\,\Gamma(8+m+n)\over \Gamma(11)}{\cal T}{}^{\a_1 \ldots\a_n}_{\b_1 \ldots\b_n},
}
where ${\cal T}{}^{\a_1 \ldots\a_n}_{\b_1 \ldots\b_n}$ are the $\g$-matrix traceless tensors
discussed in the Appendix \gammaAp\ and $\Gamma(x)$ is the gamma function.
Using ${\cal T}{}^{\a_1 \ldots\a_p}_{\a_1 \ldots\a_p} = 1$ it follows that \coefftwo
\eqn\humbps{
\int [d\l][d\lb] (\l\lb)^n e^{-(\l\lb)} = \Big({A_g\over 2\pi}\Big)^{\mkern-6mu 11}{\Gamma(8+n)\over 7!\, 60}\,.
}
For an arbitrary superfield $M(\l,\lb,\t,r)$ we define
\eqn\save{
\langle M(\l,\lb,\t,r)\rangle_{(n,g)} \equiv 
\int [d\t][dr][d\l][d\lb]\, {{\rm e}^{-(\l\lb)-(r\t)}\over (\l\lb)^{3-n}}\, M(\l,\lb,\t,r)\,,
}
which implies in particular that
\eqn\Ndef{
\langle (\l^3\t^5) \rangle_{(n,g)}
 = 2^7 R\,\Big({ 2\pi\over A_g}\Big)^{\mkern-6mu 5/2}\! \halfap{2}\, {\Gamma(8+n)\over 7!}\,\langle (\l^3\t^5)\rangle\,,
}
where $(\l^3\t^5) \equiv (\l\g^m \t)(\l\g^n \t)(\l\g^p \t)(\t\g_{mnp}\t)$ and the pure spinor bracket $\langle \ldots\rangle$ in the right-hand side
is normalized as $\langle (\l^3\t^5)\rangle = 1$ \psf.
The subscript~$g$ will be dropped whenever there is no chance for confusion.

From \opedp\ and \dalpha\ follows that
\eqn\PiPibar{
\Pi^m(z_i){\bar\Pi}^n(\bar z_j) \sim {\ap\over2}\eta^{mn}\(
2\pi\d^{(2)}(z_i-z_j) - \pi \Omega(z_i,z_j)\).
}
But using \PiPibar\ directly leads to a mixing between left- and right-movers. Instead, one can keep the two sectors separate
by expanding
$\Pi^m(z) = \hat\Pi^m(z) + \sum_{I=1}^g \Pi^m_I w_I(z)$ and computing the holomorphic square with
\eqn\cs{
\Pi^m_I \bar\Pi^n_J = -{\ap\over2} \eta^{mn}\pi\,(\Im\Omega)_{IJ}^{-1}.
}
Using this prescription, contributions containing a single $\Pi^m_I$ or
$\bar\Pi^m_I$ vanish.

We use conventions where the (anti)symmetrization over $n$ indices includes a factor of $1/n!$, the generalized
Kronecker delta is $\d^{\a_1 \ldots\a_n}_{\b_1 \ldots \b_n} \equiv \d^{[\a_1}_{\b_1} \cdots \d^{\a_n]}_{\b_n}$ and
satisfies $\d^{\a_1 \ldots\a_n}_{\a_1 \ldots \a_n} = {d\choose n}$ where $d=10$ or $d=16$ for
vector or spinor indices respectively.
The integration over $\t$ is given by $\int d^{16}\t\, \t^{\a_1} \cdots \t^{\a_{16}} = \e^{\a_1 \ldots \a_{16}}$ and
$\e^{\a_1 \ldots \a_{11}\s_1 \ldots\s_5 }\e_{\a_1 \ldots \a_{11}\b_1 \ldots\b_5} = 11!5!\,\d^{\s_1 \ldots\s_5}_{\b_1 \ldots\b_5}$.

The partition of 3-loop $d_\a$ zero-modes is denoted by $(N_1,N_2,N_3)_d$ and indicates that an expression contains
$N_I$ factors of $d^I_\a$. Furthermore, we define
\eqnn\defs
$$\eqalignno{
(\e\cdot T\cdot d^I) &\equiv \e^{\a_1 \ldots\a_{16}}T_{\a_1 \ldots\a_5}d^I_{\a_6}\cdots d^I_{\a_{16}}\,,\quad
(\lb r d^Id^J)\equiv  (\lb\g^{mnp}r)(d^I\g_{mnp}d^J).&\defs\cr
}$$
Two integrals frequently used in the next sections are summarized here,
\eqn\Tds{
\eqalign{
\int [dd^I](\e\cdot T\cdot d^I)\,d^I_{\a_1}d^I_{\a_2}d^I_{\a_3}d^I_{\a_4}d^I_{\a_5} &= 11!\,5!\,c_d\,T_{\a_1\a_2\a_3\a_4\a_5} \cr
\int [dd^I](\e\cdot T\cdot d^I)\,d^I_{\a_1}d^I_{\a_2}d^I_{\a_3}(d^I\g^{mnp}d^I) &= 11!\,5!\,96\,c_d\,(\l\g^{[m})_{\a_1}(\l\g^n)_{\a_2}(\l\g^{p]})_{\a_3}\,.
}}

\subsec Four-point SYM amplitude and kinematics

In \refs{\dhokerS,\coefftwo} the amplitudes in the Neveu-Schwarz (NS) sector were written
using the kinematic factor $K$ defined as
\eqnn\Kfactor
$$\eqalignno{
K &= F_1^{mn} F_2^{nm} F_3^{pq} F_4^{qp} + F_1^{mn} F_3^{nm} F_2^{pq} F_4^{qp} + F_1^{mn} F_4^{nm} F_2^{pq} F_3^{qp} &\Kfactor\cr
& - 4\big( F_1^{mn}F_2^{np}F_3^{pq}F_4^{qm} + F_1^{mn}F_3^{np}F_2^{pq}F_4^{qm} + F_1^{mn}F_2^{np}F_4^{pq}F_3^{qm}\big)
}$$
where $F_{mn} = k_m e_n - k_n e_m$ is the field-strength. Since the amplitudes in the pure spinor formalism are manifestly
supersymmetric one can rewrite $K$ as follows
\eqn\fourptSYM{
K = - 2^3\, 2880\, A^{\rm YM}_{1234}\,s_{12}s_{23}
}
where $A^{\rm YM}_{1234}$ is the ten-dimensional SYM amplitude normalized as
$A^{\rm YM}_{1234} = \langle V_1 E_{234}\rangle$ \nptSYM\ and $s_{ij} = k^i\cdot k^j$ are the
Mandelstam invariants. Furthermore $k^i\cdot k^i = 0$ is the massless condition and $k^1_m + \cdots + k^4_m = 0$ is
the momentum conservation relation. In order to keep the momentum expansion formul{\ae} of section 4 legible, we use
the following definitions \GreenOneLoopcoeff,
\eqn\sigmas{
\s_2 = \halfap2 (s_{12}^2 + s_{13}^2 + s_{14}^2), \quad \s_3 = \halfap3 (s_{12}^3 + s_{13}^3 + s_{14}^3)
}
and note that $\s_3 = 3 (\ap/2)^3s_{12}s_{13}s_{14}$.

\subsec{Riemann surfaces}

A holomorphic field with conformal weight one in a genus-$g$ Riemann surface $\Sigma_g$ can be expanded
in a basis of holomorphic one-forms as
$\phi(z) = {\hat \phi}(z) + \sum_{I=1}^g w_I(z)\phi^I$
and $\phi^I$ are the {\it zero modes} of $\phi(z)$. If $\{a_I,b_J\}$ are the generators of the $H_1(\Sigma_{g},\Bbb Z)=\Bbb Z^{2g}$
homology group, the holomorphic one-forms can be chosen such that for $I,J=1,2,\ldots,g$
\eqn\RieOmega{
\int_{a_I} \!w_J(z)\,dz = \d_{IJ}, \quad
\int_{b_I}\! w_J(z)\,dz=\Omega_{IJ},\quad
\int_{\Sigma_g}w_I\,\bar w_J\,\, d^2z  = 2\, \ImOmega_{IJ}\quad
}
where $\Omega_{IJ}$ is the symmetric period matrix with $g(g+1)/2$ complex degrees
of freedom \harris\ and $d^2z =i dz\wedge d{\bar z} = 2\,d{\rm Re}(z)d{\rm Im}(z)$. For the three-loop amplitude we define
\eqnn\Edef
$$\eqalignno{
\Delta(z_i;z_j;z_k) &\equiv \e^{IJK}w_{I}(z_i)w_J(z_j) w_K(z_k), &\Edef\cr
\Delta^m(z_i,z_j;z_k;z_l) &\equiv \e^{IJK}(\Pi w)^m_I(z_i,z_j) w_J(z_k)w_K(z_l),\cr
}$$
where $(\Pi w)^m_I(z_i,z_j) \equiv \Pi^m_I w_I(z_i) w_I(z_j),\hbox{(no sum in I)}$.
It follows that $\Delta^m(z_i,z_j;z_k;z_l)$ is symmetric in $(ij)$ and antisymmetric in $[kl]$ and satisfies
\eqn\DeltaPiId{
\Pi^m_I w_I(z_i) \Delta(z_j;z_k;z_l) = \Delta^m(z_i,z_j;z_k;z_l) + \Delta^m(z_i,z_k;z_l;z_j) + \Delta^m(z_i,z_l;z_j;z_k)\,.
}
Furthermore, the period matrix extends a lattice called the Jacobian variety \refs{\dhokerReview,\harris},
$J={\Bbb C}^g / ({\Bbb Z}^g+\Omega{\Bbb Z}^g)$,
which is invariant under the modular group $Sp(2g,{\Bbb Z})$.
And finally, we define
\eqn\short{
\int_{\Sigma_4} \equiv \int \prod_{i=1}^4 d^2z_i\,.
}

\subsubsec{Moduli space}

The moduli space ${\cal M}_g$ is defined as the space of inequivalent complex structures
on the Riemann surface $\Sigma_g$. It is well known that its complex dimension is
${\rm dim}_{\Bbb C}({\cal M}_g)=3g-3$, for $g>1$. We denote the complex coordinates on this space by $\tau_i$
for $i=1, \ldots,3g-3$.

For genus two and three the dimension of the moduli space is the same as the dimension
of the period matrices, i.e.,
$3g-3 = g(g+1)/2$ for $g=2,3$. So there is a one-to-one map
between inequivalent complex structures and inequivalent period matrices. This
means that for genus $g=2,3$ the scattering amplitude can be written in terms of the period
matrix instead of the moduli coordinates and Beltrami differentials.
This rewriting can be achieved using the identities \dhokerReview\
\eqn\mparam{
\int d^2z\, w_I(z)\,w_J(z) \mu_i(z) = {\delta \Omega_{IJ}\over \delta\tau_i},\quad
\int \prod_{j=1}^{6} d^2\tau_j \Bigl|\e_{i_1\ldots i_6}{\delta\Omega_{11}\over \delta\tau_{i_1}}\ldots{\delta\Omega_{33}\over \delta\tau_{i_6}}\Bigr|^2
=\int \prod_{I\leq J}^{3} d^2\Omega_{IJ},
}
where the Beltrami differential is given by $\mu_{iz}^{\bar z}= \p_z v_i^{\bar z}$ and
$ v_i^{\bar z}(z,\bar z)$ is a small complex structure deformation.

However the factor $\prod_{I\leq J}^{3} d^2\Omega_{IJ}$ is not invariant under the
modular transformation $Sp(6,{\Bbb Z})$. In general, the $Sp(2g,{\Bbb Z})$-invariant measure for the genus-$g$ moduli space is
\eqn\MeasMod{
d\mu_g \equiv {d^2\Omega_{IJ}\over (\det\ImOmega_{IJ})^{g+1}},
}
and this is precisely the measure that will be obtained from first principles in the next section for genus $g=3$.

The corresponding volume of the inequivalent period matrix space is given in \SiegelVol
\eqn\Volumes{
{\rm Vol}_g \equiv \int_{{\cal M}_g} d\mu_g = 2^{g^2+1}(2\pi)^{g(g+1)/2}\prod_{k=1}^{g}{(k-1)!\over (2k)!}|B_{2k}|\,,
}
where $B_{2k}$ are the Bernoulli numbers and the extra $2^{g(g+1)/2}$ factor in \Volumes\ compared to the original
formula in \SiegelVol\ is due to a different convention for $d^2\Omega_{IJ}$ (see e.g. \dhokerS). In particular,
\eqn\AllVolumes{
{\rm Vol}_1 = {2\pi\over3}, \quad {\rm Vol}_2={4\pi^3\over 3^3\,5}, \quad {\rm Vol}_3 = {2^6\pi^6\over 3^6\,5^2\,7}\,.
}

\subsec{The amplitude prescription}

The prescription to compute the multiloop $n$-point closed-string amplitude was given in \NMPS\ and
it becomes\foot{In the first version of this paper, we argued for an overall symmetry factor
$1/3$ for the three-loop amplitude. We thank Edward Witten for explaining to us that such factor was
incorrect.}.
\eqn\amplitude{
{\cal A}_3 =
\k^4 e^{4\l} \int_{{\cal M}_3}\mkern-1mu  \prod_{j=1}^6 d^2\tau_j \int_{\Sigma_4}
\left|\langle {\cal N}  (b,\mu_j)\, U^1(z_1) \ldots U^4(z_4)\rangle\right|^2\,,
}
for three loops and four points.
${\cal M}_3$ is the fundamental domain of the genus-three Riemann surface.
The $b$-ghost insertion is
\eqn\binsert{
(b,\mu_j) = {1\over 2\pi}\int d^2 y_j b_{zz}\mu_{j\,\bar z}^z, \quad j=1,\ldots ,6.
}
After the non-zero modes are integrated out using their OPEs, the pure spinor bracket
$\langle \ldots \rangle$ denotes the integration over the zero-modes\foot{The definition of the pure spinor bracket here should
not be confused with the standard zero mode integration $\langle (\l^3\t^5)\rangle =1$ of \psf. Since the context makes the distinction clear,
we chose not to distinguish the notation.}
\eqn\zeromodes{
\langle \ldots \rangle = \int [d\theta][dr][d\l][d\lb] \prod_{I=1}^3 [dd^I][ds^I][d\wb^I][dw^I]
}
and ${\cal N}$ is the BRST regulator discussed in \NMPS\ which can be written as
\eqn\regulator{
{\cal N} = \sum_{I=1}^3 e^{-(\l\lb) - (w^I\wb^I) - (r\t) + (s^Id^I)}.
}
After the integration over $[dd^I][ds^I][dw^I][d\wb^I]$ is performed, the remaining variables $\l^\a,\lb_\b,\t^\d$ and $r_\a$ have conformal weight zero
and therefore are the same ones which need to be integrated in the prescription of the tree-level amplitudes.
Using the Theorem~1 from Appendix~A all correlators at this stage of the computation reduce to pure spinor superspace expressions \PSSuperspace\
whose component expansions can be straightforwardly computed \refs{\PSS,\FORM}.
In particular, the last correlator to evaluate is a combination of the zero mode integration of tree-level pure spinor variables \Ndef\
and $x^m$,
\eqn\simpleA{
|\langle (\l^3\t^5) \rangle_{(n,g)}|^2 \eikx4
= (2\pi)^{10}\d^{(10)}(k) {\sqrt{2} \over 2^7 \pi^6}
\halfap{-1} \Big({\Gamma(8+n)\over 7!}\Big)^{\mkern-3mu 2} |\langle (\l^3\t^5)\rangle|^2\, {\cal I}(s_{ij}),
}
where $\d^{(10)}(k) \equiv \d^{(10)}(\sum_i k^m_i)$ and ${\cal I}(s_{ij})$ is the Koba--Nielsen factor
\eqn\KobaN{
{\cal I}(s_{ij}) = \exp\Big(-\sum_{i < j} s_{ij} G(z_i, z_j)\Big).
}
Given the above conventions, the space-time dimension of the closed-string $n$-point amplitude is independent of the genus;
$[{\cal A}_g] = n(2 + [\kappa])$. One can show that unitarity \ref\wipGM{H. Gomez and C.R. Mafra, work in progress.} requires
$\k^2 e^{-2\l} = (\ap/2)^{-2} \sqrt{2} 2^8\pi^7$, so $[\k] = -2$ and the amplitudes are dimensionless.

\newsec{The closed-string 3-loop amplitude}

At genus three there are $(16,16,16)_d$ zero-modes of $d_\a$ and $(11,11,11)_s$ zero-modes of $s^\a$.
The factor $e^{(d^Is^I)}$ in the regulator \regulator\ is the only source of $s^\a$ zero-modes so the integration over $[ds^I]$
brings down $(11,11,11)_d$ zero-modes,
\eqn\intS{
\int \prod_{I=1}^3 [ds^I]\, e^{-(d^Is^I)} = \halfap{6}\mkern-6mu{(2\pi)^{33/2}Z_3^{11}\over R^3 2^{18}(11!\,5!)^3(\l\lb)^9}
\prod_{I=1}^3 (\e\cdot T\cdot d^I)\,.
}
The remaining $(5,5,5)_d$ must come from the $b$-ghosts and the external vertices. Since the number of
$d_\a$ zero-modes from the external vertices and from each $b$-ghost can be at most four and two respectively, there are only two
possibilities for the $b$-ghosts: they provide $11$ or $12$ $d_\a$ zero modes. Note that these possibilities lead to integrations
over pure spinor variables which can be regularized using the original procedure of Berkovits \NMPS.

In the following we decompose the amplitude \amplitude\ according to the two different $b$-ghost sectors as
${\cal A}_3 = A_{11} + A_{12}$ and evaluate each sector in turn.

\subsec{$12$ $d_\a$ zero-modes from the $b$-ghosts}

In this sector there is no chance for OPE singularities between the $b$-ghosts and the
external vertices and therefore $(b,\mu)$ is still a well-defined measure \WittenCIA. To see this
note that if six $b$-ghosts provide twelve $d_\a$ zero modes, each one of them must pick the term
$(\lb \g^{mnp} r)(d\g_{mnp}d)/(192 (\l\lb)^2)$ in \bghost. The zero-mode part of each $(d\g_{mnp}d)(y)$ factor is
$$\eqalignno{
&(d^1\g_{mnp}d^1)w_1(y)w_1(y) + 2 (d^1\g_{mnp}d^2)w_1(y)w_2(y)
+ 2(d^1\g_{mnp}d^3)w_1(y)w_3(y)\cr
+ &(d^2\g_{mnp}d^2)w_2(y)w_2(y) + 2 (d^2\g_{mnp}d^3)w_2(y)w_3(y) + (d^3\g_{mnp}d^3)w_3(y)w_3(y),
}$$
and a short computation using \mparam\ gives,
\eqn\dozeds{
\int \prod_{j=1}^6  d^2\tau_j \Big|(b,\mu_j)\Big|^2 = c_{b_1}^2\int d^2\Omega_{IJ} \Big|
{B_{(4,4,4)}\over \llb^{12}} \Big|^2,
}
where $c_{b_1} = ({\ap\over2})^6 {2^3 \over (2^7\,3 \pi)^6}$ and
\eqn\Bfff{
B_{(4,4,4)} \equiv (\lb r d^1d^1)(\lb r d^1d^2)(\lb r d^1d^3)(\lb r d^2d^2)(\lb r d^2d^3)(\lb r d^3d^3).
}
Note that $B_{(4,4,4)}$ is totally symmetric in the zero-mode labels $(123)$.
Since $w^I$ and $\wb^I$ appear only in the regulator ${\cal N}$ their integration
is straightforward
\eqn\dablios{
\int \prod_{I=1}^3 [dw^I][d\wb^I]\, {\rm e}^{-(w^I\wb^I)} = {(\l\lb)^9\over (2\pi)^{33}}Z_3^{-22}\,.
}
Defining (if $B_{(p,q,r)}$ does not contain an index $m$ one omits it altogether on both sides)
\eqn\Dvect{
D^m_{(p+11,q+11,r+11)} \equiv \int \prod_{I=1}^3 [dd^I] (\e\cdot T\cdot d^I) B^m_{(p,q,r)}
}
and gathering the above results,
\eqn\exprds{
A_{12} =
{\sqrt{2}\,2^{-101} \kappa^4 e^{4\l}\over  \pi^{42}\, 3^{12}  (11!\,5!)^6} \halfap{24}
\int {d^2\Omega_{IJ} \over Z_3^{22}}\int_{\Sigma_4}\,
\Bigl|
\bigl\langle D_{(15,15,15)} U^1 U^2 U^3 U^4\bigr\rangle_{(-9)}\Bigr|^2  \bigl\langle
\prod_{j=1}^4 {\rm e}^{i k^j\cdot x^j}\bigr\rangle\,.
}
The only non-vanishing contribution to the integral in \exprds\ contains three $d_\a$ from the external vertices,
\eqnn\Ubb
$$\eqalignno{
U_1U_2U_3U_4 &=
 \halfap4 \bigl[(dW_{12})(dW_3)(dW_4)\,\eta_{12} + (dW_{13})(dW_2)(dW_4)\,\eta_{13} \cr
 &\,\quad\qquad + (dW_{14})(dW_2)(dW_3)\,\eta_{14} + (dW_{23})(dW_1)(dW_4)\,\eta_{23} &\Ubb\cr
 &\,\quad\qquad + (dW_{24})(dW_1)(dW_3)\,\eta_{24} +(dW_{34})(dW_1)(dW_2)\,\eta_{34}\bigr]\cr
&{}+\halfap3 \sum_{I=1}^3\Pi^m_I w_I(z_1) A^1_m (dW^2)(dW^3)(dW^4) + (1\leftrightarrow 2,3,4)\,,
}$$
where $W_{ij}$ denotes the BRST block \oneloopbb,
\eqn\Wij{
W_{ij}^\a= {1 \over 4} (\g_{pq} W_j)^\a  {\cal F}_i^{pq}
+  (k_j \cdot A_i)  W_j^\a  - (i \leftrightarrow j)\,.
}
Since now only the zero modes contribute, each $d_\a(z_i)$ becomes $d^I_\a w_I(z_i)$. Note that
$D_{(15,15,15)} d^I_\a d_\b^J d^K_\g w_I(z_i) w_J(z_j) w_K(z_k) = D_{(15,15,15)}d^1_\a d^2_\b d^3_\g \Delta(z_i;z_j;z_k)$ because
the only non-vanishing contribution has $(1,1,1)_d$ zero-modes and $\prod_{I=1}^3 [dd^I] (\e\cdot T\cdot d^I)$ is totally
antisymmetric in the zero-mode labels $[123]$.
Thus,
\eqnn\Mijcalc
$$\eqalignno{
D_{(15,15,15)} (dW_{ij})(dW_k)(dW_l) &= (11!\,5!\mkern2mu)^3 96^3\,c_d^3\, T_{ij,k,l}(\l,\lb,r) \Delta(z_j;z_k;z_l)\cr
D_{(15,15,15)} (dW_{i})(dW_j)(dW_k) A^m_l &= (11!\,5!\mkern2mu)^3 96^3\,c_d^3\, L^m_{ijkl}(\l,\lb,r) \Delta(z_j;z_k;z_l)&\Mijcalc\cr
}$$
where
\eqnn\Tijdef
$$\eqalignno{
T_{ij,k,l}(\l,\lb,r) &= (\lb\g^{abc}r)(\lb\g^{def}r)(\lb\g^{ghi}r)(\lb\g^{mnp}r)(\lb\g^{qrs}r)(\lb\g^{tuv}r)\cr
 &\quad\times(\l\g^{adefm}\l)(\l\g^{bghit}\l)(\l\g^{uqrsn}\l)(\l\g^c W_{ij})(\l\g^p W_k)(\l\g^v W_l)\,,\cr
L_{ijkl}^m(\l,\lb,r) &= (\lb\g^{abc}r)(\lb\g^{def}r)(\lb\g^{ghi}r)(\lb\g^{mnp}r)(\lb\g^{qrs}r)(\lb\g^{tuv}r)&\Tijdef\cr
 &\quad\times(\l\g^{adefm}\l)(\l\g^{bghit}\l)(\l\g^{uqrsn}\l)(\l\g^c W_i)(\l\g^p W_j)(\l\g^v W_k)A^m_l.
}$$
As shown in the Appendix \gammaAp, it is always possible to rewrite the $\lb^n\l^{n+3}$ dependence in \Tijdef\ as $\llb^n \l^3$ when performing the
zero mode integrals and therefore we write
$T_{ij,k,l}(\l,\lb,r) = \llb^6 T_{ij,k,l}(\l,r)$ and drop the $(\l,r)$ arguments from now on. Note that $T_{ij,k,l}$ is antisymmetric in $[ij]$ and $[kl]$ and
$L_{ijkl}^m$ is antisymmetric in $[ijk]$.

Plugging the above results in \exprds\ and using the identity \DeltaPiId\ together with
the definitions $M_{ij,k,l} = s_{ij}^{-1}T_{ij,k,l}$, $X_{ij} = (\ap/2)s_{ij}\eta_{ij}$ yields
\eqnn\AtwelveFinal
$$\eqalignno{
A_{12} &= {\sqrt{2}\pi^6\over 2^{23}\,3^6} \kappa^4 e^{4\l}\halfap6
\int {d^2\Omega_{IJ} \over Z_3^{-10}} \int_{\Sigma_4}
\langle|{\cal K}_{12}|^2  \bigr\rangle_{(-3)}  \bigl\langle
\prod_{j=1}^4 {\rm e}^{i k^j\cdot x^j}\bigr\rangle &\AtwelveFinal
}$$
where
\eqnn\Ktwelve
$$\eqalignno{
{\cal K}_{12} &= M_{12,3,4}\Delta(z_2;z_3;z_4)\,X_{12} + M_{13,2,4}\Delta(z_3;z_2;z_4)\,X_{13}
+ M_{14,2,3}\Delta(z_4;z_2;z_3)\,X_{14}\cr 
&+ M_{23,1,4}\Delta(z_3;z_1;z_4)\,X_{23} + M_{24,1,3}\Delta(z_4;z_1;z_3)\,X_{24} + M_{34,1,2}\Delta(z_4;z_1;z_2)\,X_{34}\cr
&+ \Delta^m(z_1,z_2;z_3;z_4)\big[ L^m_{1342} + L^m_{2341}\big] + \Delta^m(z_1,z_3;z_2;z_4)\big[ L^m_{1243} + L^m_{3241}\big] \cr
& + \Delta^m(z_1,z_4;z_2;z_3)\big[ L^m_{1234} + L^m_{4231}\big] + \Delta^m(z_2,z_3;z_1;z_4)\big[ L^m_{2143} + L^m_{3142}\big]\cr
& + \Delta^m(z_2,z_4;z_1;z_3)\big[ L^m_{2134} + L^m_{4132}\big] + \Delta^m(z_3,z_4;z_1;z_2)\big[ L^m_{3124} + L^m_{4123}\big]. &\Ktwelve\cr
}$$

\subsec{$11$ $d_\a$ zero-modes from the $b$-ghosts}

By not using the Siegel-gauge vertex operators of \yuri\ one could in principle face problems with
the consistency condition for $(b,\mu)$ discussed in \WittenCIA.
However, in the low-energy limit discussed here, this
potential complication can be ignored since the only contribution comes
from terms in which the b ghost does not have singular OPEs\foot{We are grateful to Nathan Berkovits for discussions and for his
comments on the draft at this point.}.
To see this note that one possibility to obtain $11$ $d_\a$ zero modes
out of six $b$-ghosts is given by
\eqn\problem{
b^6  \longrightarrow {(\l\lb)^{-12} \over (192)^6 }\halfap6 \big[  (\lb\g^{mnp}r)(d\g_{mnp}d)\big]^6\,,
}
where one of the $d_\a(y_i)$ is the non-zero-mode part $\hat d_\a(y_i)$ and contracts through the OPE \opedp\ with an external vertex ($y_i$ denotes
the position of the corresponding $b$-ghost). However this term is of order $D^8 R^4$ and there are no inverse powers of Mandelstam invariants
coming from the integration over the vertex positions since there are no simple pole singularities among them \BerkovitsAW, as they must contribute the
four remaining $d_\a$ zero modes. The claim that \problem\ leads to terms of order $D^8R^4$ is easy to verify. The external
vertices contribute $W^4$ superfields, the OPE between
$\hat d_\a(y)$ and one superfield $W^\b$ gives $D_\a W^\b$ and each $r_\a$ from \problem\ counts as a covariant
derivative $D_\a$ because of the factor $e^{-(r\t)}$ in the regulator $\cal N$. This gives kinematic terms proportional to
$\langle D_\a^7 W^4\rangle = \langle k^3 W^3\cF\rangle = k^4 F_{mn}^4$ whose holomorphic square is $D^8R^4$. The other possibilities
of $b$-ghost singularities are similarly analyzed. Therefore the
terms which might be affected by the issues pointed out in \WittenCIA\ 
do not affect the leading order terms $D^6R^4$ and will not be considered in the following.

When the $b$-ghosts have no singularities with the vertices and contribute $11$ zero-modes of $d_\a$, one possibility is
\eqn\zeroafter{
b^6  \longrightarrow
{\llb^{-13} \over 16(192)^5 }\halfap6 (r\g_{qrs}r)(\lb \g^{q}d)N^{rs} \big[  (\lb\g^{mnp}r)(d\g_{mnp}d)\big]^5,
}
but it vanishes upon integration over $[dw]$ because $\int[dw][d\wb] w_\a e^{- (w^I \wb_I)} = 0$.
The other possibility is
\eqn\opeb{
b^6  \longrightarrow {(\l\lb)^{-11} \over 2(192)^5 }\halfap5 (\lb \g_{q}d)\Pi^{q} \big[  (\lb\g^{mnp}r)(d\g_{mnp}d)\big]^5
}
where the $\Pi^m(y)$ field is proportional to its zero modes $\Pi_m^I w_I(y)$. In this case the integration over the positions of the
$b$-ghosts can be carried out,
\eqn\onzeds{
\int \prod_{j=1}^6  d^2\tau_j \Big|(b,\mu_j)\Big|^2 =
c_{b_2}^2 \int d^2\Omega_{IJ}\Bigl| {1\over (\l\lb)^{11}}\big(
\Pi_m^1 B^m_{(3,4,4)} + \Pi_m^2 B^m_{(4,3,4)} + \Pi_m^3 B^m_{(4,4,3)}\big)\Bigr|^2
}
where $c_{b_2} = {2^2(\ap\mkern-2mu/2)^5\over 4\pi\,(2^7\,3\pi)^5} = 48(\ap/2)^{-1}\, c_{b_1}$ and
\eqnn\Beleven
$$\eqalignno{
B^m_{(3,4,4)}  ={}& +2(\lb \g^m d^1) (\lb r d^1d^2) (\lb r d^1d^3) (\lb r d^2d^2) (\lb r d^2d^3)(\lb r d^3d^3)\cr
& - (\lb \g^m d^2) (\lb r d^1d^1) (\lb r d^1d^3) (\lb r d^2d^2) (\lb r d^2d^3) (\lb r d^3d^3)\cr
& + (\lb \g^m d^3 ) (\lb r d^1d^1) (\lb r d^1d^2) (\lb r d^2d^2) (\lb r d^2d^3) (\lb r d^3d^3)\,,&\Beleven\cr
}$$
while $B^m_{(4,3,4)}$ and $B^m_{(4,4,3)}$ are obtained from $B^m_{(3,4,4)}$ by swapping $d^1_\a\leftrightarrow d^2_\a$ and $d^1_\a\leftrightarrow d^3_\a$, respectively.
Furthermore, note that $B^m_{(3,4,4)}$ is symmetric under $d^2_\a\leftrightarrow d^3_\a$.

Taking into account that $c_{b_2} = 48(\ap/2)^{-1}\, c_{b_1}$ and using the definition \Dvect\ leads to the following expression
for $A_{11}$,
\eqnn\Aeleven
$$\eqalignno{
A_{11} &=
{\sqrt{2}\,2^{-93} \kappa^4 e^{4\l}\over  \pi^{42}\, 3^{10}  (11!\,5!)^6} \halfap{22}
\int {d^2\Omega_{IJ} \over Z_3^{22}} \int_{\Sigma_4}\, \,\eikx4 &\Aeleven\cr
&\quad \times \Bigl|
\bigl\langle (\Pi_m^1 D^m_{(14,15,15)} + \Pi_m^2 D^m_{(15,14,15)} + \Pi_m^3 D^m_{(15,15,14)}) U^1 U^2 U^3 U^4\bigr\rangle_{(-8)}\Bigr|^2.
}$$
Each external vertex $U^i$ contribute through the term $(\ap/2)(dW^i)(z_i)$
and the integration over the $d$ zero-modes can be carried out by using the following formulae,
\eqnn\Sdefin
$$\eqalignno{
D^m_{(14,15,15)} (d^1W^1)(d^1W^2)(d^2W^3)(d^3W^4) &= + (11!\,5!)^3 96^2 c_d^3\, S^m_{1234}(\l,\lb,r)\cr
D^m_{(15,14,15)} (d^2W^1)(d^2W^2)(d^1W^3)(d^3W^4) &= - (11!\,5!)^3 96^2 c_d^3\, S^m_{1234}(\l,\lb,r)&\Sdefin\cr
D^m_{(15,15,14)} (d^3W^1)(d^3W^2)(d^1W^3)(d^2W^4) &= + (11!\,5!)^3 96^2 c_d^3\, S^m_{1234}(\l,\lb,r)\cr
}$$
where
\eqn\Sdef{
S^m_{1234}(\l,\lb,r) = S^{(1)\,m}_{1234}(\l,\lb,r) + S^{(2)\,m}_{1234}(\l,\lb,r) - S^{(2)\,m}_{1243}(\l,\lb,r)
}
and
\eqnn\Sthree
$$\eqalignno{
S^{(1)\,m}_{1234}(\l,\lb,r) ={}& 2\,(\lb\g^m \g^{a_1}\l)(\lb\g_{m_1 n_1 p_1} r) (\lb\g_{m_2 n_2 p_2} r) (\lb\g_{m_3 n_3 p_3} r)(\lb\g_{m_4 n_4 p_4} r)(\lb\g_{m_5 n_5 p_5} r)\cr
&\times (\l\g^{a_2m_1n_1p_1m_3}\l)(\l\g^{a_3m_2n_2p_2m_5}\l)(\l\g^{n_3m_4n_4p_4n_5}\l)\cr
&\times (W^1\g^{a_1a_2a_3}W^2)(\l\g^{p_3}W^3)(\l\g^{p_5}W^4)\cr
S^{(2)\,m}_{1234}(\l,\lb,r) ={}& 96\,(\lb\g^m \g^{m_3}\l)(\lb\g_{m_1 n_1 p_1} r) (\lb\g_{m_2 n_2 p_2} r) (\lb\g_{m_3 n_3 p_3} r)(\lb\g_{m_4 n_4 p_4} r)(\lb\g_{m_5 n_5 p_5} r)\cr
&\times (\l\g^{m_1m_2n_2p_2m_5}\l)(\l\g^{n_3m_4n_4p_4n_5}\l)\cr
&\times (\l\g^{n_1}W^1)(\l\g^{p_1}W^2)(\l\g^{p_3}W^3)(\l\g^{p_5}W^4)\,. &\Sthree\cr
%
}$$
Note from the definition \Sdefin\ that $S^m_{1234}(\l,\lb,r)$ is symmetric in the particle labels $(12)$ and antisymmetric in $[34]$.
The explicit expression for $S^{(1)}_{1234}(\l,\lb,r)$ is symmetric in $(12)$ and antisymmetric in $[34]$ whereas $S^{(2)}_{1234}(\l,\lb,r)$ is symmetric in $(12)$,
so \Sdef\ indeed has the required symmetries. According to the procedure of Appendix~A we write
$S^m_{1234}(\l,\lb,r) = \llb^6 S^m_{1234}(\l,r)$ and drop the arguments $(\l,r)$ in the following.

After expanding $d_\a(z) = \hat d_\a(z) + d^I_\a w_I(z)$ and using
\Sdefin\ together with the symmetry properties of $S^m_{1234}$ it is a matter of bookkeeping the permutations to arrive at
\eqn\Pims{
(\Pi_m^1 D^m_{(14,15,15)} + \Pi_m^2 D^m_{(15,14,15)} + \Pi_m^3 D^m_{(15,15,14)}) U^1 U^2 U^3 U^4 =
\halfap4 (11!5!)^3\,96^2\,c_d^3\,\llb^6 {\cal K}_{11}
}
where
\eqnn\Keleven
$$\eqalignno{
{\cal K}_{11} =
&{}+ S_{1234}^m \DeltaPi1,2,3,4,m + S_{1324}^m \DeltaPi1,3,2,4,m + S_{1423}^m \DeltaPi1,4,2,3,m \cr
&{}+ S_{2314}^m \DeltaPi2,3,1,4,m + S_{2413}^m \DeltaPi2,4,1,3,m + S_{3412}^m \DeltaPi3,4,1,2,m\,.\cr
& &\Keleven
}$$
Finally,
\eqn\AelevenFinal{
A_{11}= \kappa^4 e^{4\l}\,{\sqrt{2}\pi^6 \over 2^{25}3^6}\,
\halfap6\mkern-4mu \int {d^2\Omega_{IJ} \over Z_3^{-10}}\int_{\Sigma_4}\, \langle
\big|{\cal K}_{11} \big|^2\rangle_{(-2)}\eikx4.
}
Therefore from \AtwelveFinal\ and \AelevenFinal\ the three-loop amplitude ${\cal A}_3 = A_{12} + A_{11}$ becomes
\eqn\Athree{
{\cal A}_3 = \kappa^4 e^{4\l}\,{\sqrt{2}\pi^6 \over 2^{23}3^6}\,
\halfap6\!\! \int_{{\cal M}_3}\! {d^2\Omega_{IJ} \over (\det(2\ImOmega))^5}\int_{\Sigma_4}\Big[
\langle |{\cal F}|^2\rangle_{(-3)} + \langle|{\cal T}|^2\rangle_{(-3)}\Big]\eikx4
}
where \Zg\ has been used,
\eqnn\calTdef
$$\eqalignno{
{\cal F} = &{}+ M_{12,3,4}\Delta(z_2;z_3;z_4)\,X_{12} + M_{13,2,4}\Delta(z_3;z_2;z_4)\,X_{13}
+ M_{14,2,3}\Delta(z_4;z_2;z_3)\,X_{14}\cr
&+ M_{23,1,4}\Delta(z_3;z_1;z_4)\,X_{23} + M_{24,1,3}\Delta(z_4;z_1;z_3)\,X_{24} + M_{34,1,2}\Delta(z_4;z_1;z_2)\,X_{34}\cr
{\cal T} =
&{}+ T_{1234}^m \DeltaPi1,2,3,4,m + T_{1324}^m \DeltaPi1,3,2,4,m + T_{1423}^m \DeltaPi1,4,2,3,m \cr
&{}+ T_{2314}^m \DeltaPi2,3,1,4,m + T_{2413}^m \DeltaPi2,4,1,3,m + T_{3412}^m \DeltaPi3,4,1,2,m&\calTdef\cr
}$$
and (the other $T^m_{ijkl}$ follow from relabeling),
\eqn\Tmdef{
T^m_{1234} = L_{1342}^m + L_{2341}^m + {5\over 2} S_{1234}^m\,.
}
The factor $5$ in \Tmdef\ is due to $\langle \ldots\rangle_{(-2)} = 5\langle \ldots\rangle_{(-3)}$ and follows from \Ndef.
The factor $1/2$ accounts for the different overall normalizations of \AelevenFinal\ and \AtwelveFinal.

After using \simpleA\ the three-loop amplitude \Athree\ becomes
\eqn\AthreeFinal{
{\cal A}_3 = (2\pi)^{10}\d^{(10)}(k){\kappa^4 e^{4\l}\over 2^{31}3^8\,5^2\,7^2}\,
\halfap5\!\! \int_{{\cal M}_3}\! {d^2\Omega_{IJ} \over (\det(2\ImOmega))^5}\int_{\Sigma_4}\Bigl[
\langle |{\cal F}|^2\rangle + \langle|{\cal T}|^2\rangle\Bigr]\,{\cal I}(s_{ij})\,.
}

\subsec{The low-energy limit $D^6R^4$}

Since the superfields in ${\cal F}$ and ${\cal T}$ have component expansions
terms of order $k^4 F^4_{mn}$ and $k^3 F^4_{mn}$ one might naively
expect that only $\langle|{\cal T}|^2\rangle$ in \Athree\ contributes to the low-energy limit of order
$D^6 R^4$. However some integrals in $\langle|{\cal F}|^2\rangle$ contain kinematic poles which reduce
its contribution from $D^8R^4$ to $D^6 R^4$. In fact, the
integration by parts identities of Appendix~\IBPappendix\ show that\foot{In the first version of this
paper the result of the integral $\int_{\Sigma_4}
\Omega_{12}\Delta(z_2;z_3;z_4)\bar\Delta(z_1;z_3;z_4)$ was incorrectly stated as $36\det(2\ImOmega)$ instead of
$12\det(2\ImOmega)$. Using a notation where
$X_{IJ}\equiv \Im\Omega_{IJ}$, the definitions \PiOPE\ and \Edef\ imply that the integral is 
$$\eqalignno{
&= \int_{\Sigma_4} w_I(z_1)X^{-1}_{IJ}\bar w_J(\bar z_2) \e^{KLM}w_K(z_2)w_L(z_3)w_M(z_4) \e^{PQR}\bar w_P(\bar
z_1)\bar w_Q(\bar z_3)\bar w_R(\bar z_4)\cr
&= 2^4  X_{IP}X_{IJ}^{-1} X_{JK} \e^{KLM}\e^{PQR}X_{LQ}X_{MR} = 2^4 X_{PK}X_{QL}X_{RM}
\e^{KLM}\e^{PQR}
 = 12 \det(2\Im\Omega)
}$$
where $X_{IJ}^{-1}X_{JK} = \d_{IK}$, $X_{PK}X_{QL}X_{RM} \e^{KLM}\e^{PQR}= 3!\det(X)$ and $\det(2X) = 2^3 \det(X)$ have been used.
A similar mistake in the calculation of $\int_{\Sigma_4}\! |{\cal T}|^2$ led to an incorrect overall
coefficient for the low-energy limit which was bigger by a factor of 3.}
\eqnn\leadingF
$$\eqalignno{
\int_{\Sigma_4}\langle|{\cal F}|^2\rangle{\cal I}(s_{ij}) &= - \pi\halfap{}\,\langle{\cal K}\rangle \int_{\Sigma_4} \Omega_{12}\Delta(z_2;z_3;z_4)\bar\Delta(z_1;z_3;z_4)
 + {\cal O}(\ap^2)\cr
&= - 12\pi\halfap{}\,\langle{\cal K}\rangle \det(2\ImOmega) + {\cal O}(\ap^2) &\leadingF\cr
}$$
where
\eqn\Kfinal{
{\cal K} =   {|T_{23,1,4}|^2\over s_{23}} + {|T_{24,1,3}|^2\over s_{24}} + {|T_{34,1,2}|^2\over s_{34}}
   + {|T_{12,3,4}|^2\over s_{12}} + {|T_{13,2,4}|^2\over s_{13}} + {|T_{14,2,3}|^2\over s_{14}}\,
}
is such that $\langle {\cal K}\rangle$ is also of order $D^6R^4$.

To compute the low-energy limit of $\int_{\Sigma_4}\! |{\cal T}|^2\,{\cal I}(s_{ij})$ it will
be convenient to use the symmetry relations\foot{We thank Piotr Tourkine for pointing out the first
symmetry relation in \piotr\ and
Oliver Schlotterer for emphasizing its role in simplifying the expression of ${\cal T}$ and
${\cal L}\cdot \tilde{\cal L}$.}
\eqn\piotr{
T^m_{1234} + T^m_{3124} + T^m_{2314} = 0,\quad T^m_{1234} = T^m_{2134},\quad T^m_{1234} = - T^m_{1243}
}
to eliminate $T^m_{1234}$, $T^m_{1324}$ and $T^m_{2314}$ from \calTdef\ (using, for instance,
$T^m_{1234} = T^m_{1423} + T^m_{2413}$). Doing this and applying \DeltaPiId\ one arrives at
\eqn\simpleT{
{\cal T} = 
T^m_{1423} \sum_{I=1}^3\Pi_I^m w_I(z_1)\Delta_{234}
+ T^m_{2413} \sum_{I=1}^3\Pi_I^m w_I(z_2)\Delta_{134}
+ T^m_{3412} \sum_{I=1}^3\Pi_I^m w_I(z_3)\Delta_{124}\,.
}
After setting ${\cal I}(s_{ij})=1$ (since there are no kinematic poles),
the contribution of $\int_{\Sigma_4}\langle |T|^2\rangle$ to the low-energy limit require
the following integrals
\eqnn\Mintegrals
$$\eqalignno{
\int_{\Sigma_4} \Pi^m_I\bar\Pi^n_J w_I(z_1)\bar w_J(\bar z_1) \Delta_{234}\bar\Delta_{234} &= -
36\pi\eta^{mn}\halfap{}\det(2\Im\Omega) &\Mintegrals\cr
\int_{\Sigma_4} \Pi^m_I\bar\Pi^n_J w_I(z_1)\bar w_J(\bar z_2) \Delta_{234}\bar\Delta_{341} &= -
12\pi\eta^{mn}\halfap{}\det(2\Im\Omega)\,.\cr
}$$
and yields
\eqn\calTint{
\int_{\Sigma_4}\! |{\cal T}|^2 = -12\pi\halfap{}\det(2\Im\Omega){\cal L}\cdot\tilde{\cal L}
}
where
\eqn\calLcalL{
{\cal L}\cdot\tilde{\cal L}\equiv |T^m_{1234}|^2
+ |T_{1324}^m|^2 + |T_{1423}^m|^2 + |T_{2314}^m|^2 + |T_{2413}^m|^2 + |T_{3412}^m|^2\,.
}
Therefore the low-energy limit of the three-loop amplitude \AthreeFinal\ is given by
\eqnn\LEthreeFirst
$$\eqalignno{
{\cal A}_3 &= -(2\pi)^{10}\d^{(10)}(k)\,\halfap6\, \bigl\langle {\cal K} + {\cal L}\cdot\tilde{\cal L}  \bigr\rangle\, \kappa^4 e^{4\l}\,
{\pi\over 2^{29}3^7\,5^2\,7^2}\,
\int_{{\cal M}_3} {d^2\Omega_{IJ} \over (\det(2\ImOmega))^4}\cr
&= -(2\pi)^{10}\d^{(10)}(k)\,\halfap6 \bigl\langle {\cal K} + {\cal L}\cdot\tilde{\cal L}  \bigr\rangle\,\kappa^4 e^{4\l}\,
{\pi \zeta_6 \over 2^{35}\, 3^{10}\, 5^3\, 7^2} &\LEthreeFirst
}$$
where the integral is $2^{-12}\,{\rm Vol}_3$ and we used $\zeta_6 = \pi^6/945$.
A long calculation gives \PSS
\eqn\LongTime{
\bigl\langle {\cal K} + {\cal L}\cdot\tilde{\cal L}  \bigr\rangle =
-2^{35}\, 3^7\, 5^3\, 7^2\,(s_{12}^3 + s_{13}^3 + s_{14}^3)\, K{\bar K}\,
}
irrespective of whether it is for type IIA or IIB, confirming the theorem of \NathanTheorem.
Therefore
\eqn\LEthree{
{\cal A}_3 = (2\pi)^{10}\delta^{(10)}(k)\, \k^4e^{4\l}\, {\pi\,\zeta_6\over 3^3}\halfap6 (s_{12}^3 + s_{13}^3 + s_{14}^3)\,K{\bar K}
}
is the low-energy limit of the type IIA and IIB three-loop amplitude.

\newsec Perturbative calculations versus S-duality predictions

We first review the one- and two-loop comparisons between S-duality predictions and perturbative
amplitude calculations of \refs{\GreenOneLoopcoeff,\dhokerS} using our conventions. After that we
extend their analysis to include the three-loop result \LEthree. We will find that the amplitude we computed in \LEthree\ agrees with
the prediction of Green and Vanhove \Greenthreeloop.

\subsec One- and two-loops

The closed-string massless four-point amplitudes at genus $0$, $1$ and $2$ computed in \coefftwo\ (including their overall coefficients) are
given by (see also \dhokerS),
\eqnn\AmpCoeffs
$$\eqalignno{
{\cal A}_0 &= (2\pi)^{10}\d^{(10)}(k)\, \halfap3  K{\bar K}\, \k^4 e^{-2\l}\, {\sqrt{2} \over 2^{16} \pi^5} \, {\cal B}_0(s_{ij}) &\AmpCoeffs\cr
{\cal A}_1 &= (2\pi)^{10}\d^{(10)}(k)\, \halfap3  K{\bar K}\, \k^{4}\,{1 \over 2^{14} \pi^2}
\int_{{\cal M}_1}\mkern-3mu {d^{2}\tau \over \tau_2^2}{\cal B}_1(s_{ij}|\tau)\cr
{\cal A}_2 &=
(2 \pi)^{10}\d^{(10)}(k)\, \halfap3  K{\bar K}\, \kappa^4 e^{2\l}\, {\sqrt{2} \over 2^{15}}
 \int_{{\cal M}_2}\! {d^2\Omega_{IJ} \over (\det(\Im\Omega))^3}
 {\cal B}_2(s_{ij}|\Omega)
}$$
where\foot{${\cal B}_0(s_{ij})$ here is ${1\over 2\pi}C(s,t,u)$ from \coefftwo. Furthermore,
see \GreenDhoker\ for
a recent attempt to evaluate non-leading terms at two-loops and \SS\ for an elegant way to rewrite the tree-level expansion.}
\refs{\GreenOneLoopcoeff,\dhokerS},
\eqnn\GammaExp
$$\eqalignno{
{\cal B}_0(s_{ij}) &= {\Gamma(-\ap s_{12}/2)\Gamma(-\ap s_{13}/2) \Gamma(- \ap s_{14}/2) \over \Gamma(1+ \ap s_{12}/2)\Gamma(1+ \ap
s_{13}/2)\Gamma(1+\ap s_{14}/2)} &\GammaExp \cr
&= {3\over \s_3} + 2\zeta_3 + \zeta_5 \s_2 + {2\over 3}\zeta_3^2 \s_3 + \cdots \cr
{\cal B}_1(s_{ij}|\tau) &=  \int \prod_{i=2}^4 {d^2 z_i\over \tau_2}\, {\cal I}(s_{ij}) = 2^3\Bigl(1+ {\zeta_3\over 3}\s_3 + \cdots\cr
{\cal B}_2(s_{ij}|\Omega) &=  \int_{\Sigma^4} { |{\cal Y}|^2 \over (\det(\Im\Omega))^2}\, {\cal I}(s_{ij}) = 2^7 \s_2 + \cdots 
}$$
Plugging in the volume of the moduli spaces \AllVolumes\ one obtains the following low-energy expansions,
\eqnn\leadingAmpsTree
\eqnn\leadingAmpsOne
\eqnn\leadingAmpsTwo
$$\eqalignno{
{\cal A}_0 &= (2\pi)^{10}\d^{(10)}(k)\,\halfap3 K{\bar K}\, \k^{4}e^{-2\l} \,{\sqrt{2} \over 2^{16} \pi^5} \,
\Bigl[ {3\over \s_3} + 2\zeta_3 + \zeta_5 \s_2 + {2\over 3}\zeta_3^2 \s_3 + \cdots &\leadingAmpsTree\cr
{\cal A}_1 &= (2\pi)^{10}\d^{(10)}(k)\,\halfap3 K{\bar K}\, \k^{4}\,{1  \over 2^{10}\,3\pi} \Big[1+ {\zeta_3 \over 3}\s_3 +
\cdots &\leadingAmpsOne\cr
{\cal A}_2 &= (2\pi)^{10}\d^{(10)}(k)\,\halfap3 K{\bar K}\, \k^{4}e^{2\l}\,{\sqrt{2}\pi^3  \over 2^{6}\,3^3\,5}\, \Bigl[ \s_2 + \cdots
&\leadingAmpsTwo\cr
}$$
The $SL(2,\Bbb Z)$-duality predictions for the perturbative effective action are \refs{\GGRq,\GreenKVan,\Greenthreeloop}
\eqnn\SRq
\eqnn\SDqRq
\eqnn\SDsRq
$$\eqalignno{
S^{\ap^3} &= C_1 \int d^{10}x\sqrt{-g}\, {\cal R}^4(2\zeta_3 e^{-2\phi} + {2\pi^2\over 3})\,, &\SRq\cr
S^{\ap^5} &= C_2\int d^{10}x\sqrt{-g}\, D^4{\cal R}^4(2\zeta_5 e^{-2\phi} + {8\over 3}\zeta_4e^{2\phi})\,, &\SDqRq\cr
S^{\ap^6} &= C_3\int d^{10}x\sqrt{-g}\, D^6{\cal R}^4(4\zeta_3^2 e^{-2\phi} + 8\zeta_2\zeta_3 + {48\over 5}\zeta_2^2 e^{2\phi} + {8\over 9}\zeta_6 e^{4\phi}).&\SDsRq\cr
}$$
where the precise definitions of $R^4$, $D^4{\cal R}^4$ and $D^6{\cal R}^4$ and the constants
$C_{\{1,2,3\}}$ will not be needed in the following discussion since only the ratios of the interactions at 
different loop orders will be important.

Matching the ratios of the $\ap^3$ interactions at one-loop and tree-level leads to a relation between $e^\phi$ and $e^\l$,
\eqn\ratioAmps{
 {\sqrt{2}2^4\pi^4 e^{2\l}\over 3\zeta_3} = {e^{2\phi}\pi^2\over 3\zeta_3}\quad\rightarrow\quad e^{2\phi} = e^{2\l} \sqrt{2} 2^4 \pi^2,
}
where the left-hand side follows from the amplitudes while the right-hand side from the effective action \SRq. Now
one can compare the S-duality predictions for the amplitudes at order $\ap^5$ and $\ap^6$ (denoted with a Latin capital $A^{\ap^n}$) and the perturbative results.

For the $\ap^5$ interaction, the ratio between the two-loop and tree-level interactions in the effective action \SDqRq\ is
${4\zeta_4 \over 3\zeta_5}e^{4\phi}$ and leads to the prediction
\eqn\ratioTwoAmp{
A_2^{\ap^5} = {\cal A}_0^{\ap^5}{2^{11}\pi^4 \zeta_4\over 3\zeta_5}e^{4\l}
=(2\pi)^{10}\d^{(10)}(k)\, \k^{4}e^{2\l}\,\halfap3 K{\bar K}\, {\sqrt{2}\zeta_4  \over 2^{5}\,3\,\pi}\, \s_2\,,
}
which agrees with the two-loop perturbative calculation \leadingAmpsTwo\ (recall that $\zeta_4 = \pi^4/90$).

For the $\ap^6$ interaction, the ratio between the one-loop and tree-level terms following from the effective action \SDsRq\ is
$2\zeta_2/\zeta_3 e^{2\phi} = \zeta_2/\zeta_3\sqrt{2}2^5\pi^2e^{2\l}$
and implies
\eqn\DsixOneloop{
A_1^{\ap^6} = (2\pi)^{10}\delta^{(10)}(k)\, \halfap3 K{\bar K}\, \k^4\,  {\zeta_2\zeta_3\over 2^9 3\pi^3 }\, \s_3\,,
}
which agrees with the one-loop perturbative calculation \leadingAmpsOne, in accord with the analysis of \GreenOneLoopcoeff.

\subsec Three-loops

Similarly, the ratio between the
three-loop and tree-level terms of \SDsRq\
$$
{2\zeta_6\over 9\zeta_3^2}\,e^{6\phi} = {\sqrt{2}2^{14}\pi^6\zeta_6\over 9\zeta_3^2}\,e^{6\l} = {{\cal A}^{\ap^6}_3\over {\cal A}^{\ap^6}_0}
$$
predicts the following three-loop amplitude
\eqn\threeloopPrediction{
A_3^{\ap^6} =
(2\pi)^{10}\delta^{(10)}(k)\,\halfap3 K{\bar K}\, \k^4e^{4\l}\, {\pi\zeta_6\over 3^3}\,\s_3\,,
}
in complete agreement with the first principles perturbative calculation \LEthree.

\bigskip
\noindent{\bf Acknowledgements:} CRM thanks Michael Green for many discussions over the years and
for his continuous interest. We also thank Edward Witten for making it clear that no symmetry
argument could account for an overall factor of $1/3$ as argued in the first version of this paper.
CRM also thanks Oliver Schlotterer for collaboration on related topics.
We would like to thank Nathan Berkovits for discussions and to Oscar A. Bedoya for
reading the manuscript. H.G thanks Freddy Cachazo for his comments and discussions about Riemann surfaces
and Andrei Mikhailov for discussions about the pure spinor formalism.
H.G is grateful to the Albert-Einstein-Institut,
DAMTP, IFT-UNESP and especially to the Perimeter Institute for Theoretical Physics
for warm hospitality during stages of this work.
CRM thanks the Albert-Einstein-Institut in Potsdam for the wonderful work environment during the early stages of this work.
The work of H.G is
supported by FAPESP grant 2011/13013-8, 13/11409-7 and CRM acknowledges support by the European Research Council Advanced
Grant No. 247252 of Michael Green.

\appendix{A}{A general formula for integration of pure spinors}
\applab\gammaAp

\noindent
Component expansions of a general ghost-number-three pure spinor superspace expression of the
form $\la1 \ldots \la{n+3} \lb_{\b_1} \ldots\lb_{\b_n}f_{\a_1 \ldots\a_{n+3}}^{\b_1 \ldots\b_n}(\t,r)$ can be
computed most conveniently by first rewriting it in the form
$\llb^n \la1 \ldots \la3 f_{\a_1 \ldots\a_3}(\t,r)$. Doing that allows the straightforward application of the formula \Ndef\
and the identities listed in the appendix of \anomalia.
The case  $n=1$ was discussed in \fiveone, now the solution for general $n$ will be presented.

Let ${\cal T}{}^{\a_1 \ldots\a_n}_{\b_1 \ldots\b_n}$ denote a $SO(10)$-invariant tensor which is
symmetric and $\g$-traceless in both sets of indices. When $n$ is even there is $(n/2+1)$-dimensional basis \HoogeveenHK,
\eqn\Tndef{
{\cal T}{}^{\a_1 \ldots\a_n}_{\b_1 \ldots\b_n} = \sum_{k=0}^{n/2} c_k^{(n)} T^{(n)}_k\,,
}
$$
T_k^{(n)} = \d^{(\a_1}_{(\b_1}\cdots
\d^{\a_{n-2k}}_{\b_{n-2k}}(\g\cdot\g)^{\a_{n-2k+1}\a_{n-2k+2}}_{\b_{n-2k+1}\b_{n-2k+2}}
\cdots
(\g\cdot\g)^{\a_{n-1}\a_{n})}_{\b_{n-1}\b_{n})}\,,
$$
and $(\g\cdot\g)^{\a_1\a_2}_{\b_1\b_2} \equiv \g_m^{\a_1\a_2}\g^m_{\b_1\b_2}$. Imposing the $\g$-traceless condition leads to 
a recurrence relation for the coefficients $c_k^{(n)}$ \ref\StahnPriv{Christian Stahn, private communication in 2008} and
the normalization condition $T^{\a_1 \ldots \a_n}_{\a_1 \ldots \a_n} = 1$ relates the
coefficient $c_0^{(n)}$ with the dimension of the pure spinor representation $N_n \equiv {\rm dim}([0000n])$,
\eqn\recurr{
c_{k+1}^{(n)} = - {(n-2k)(n-2k-1)\over 8 (k+1)(n-k+2)}c_k^{(n)}\,,\quad c_0^{(n)} = 1/N_n\,,
}
where\foot{$N_n$ can be obtained from
$(1+t)(1+4t+t^2)(1-t)^{-11} = 1+ \sum_{n\ge 1} N_n t^n$ \refs{\character,\aldoPF}.}
\eqnn\dimN
$$\eqalignno{
N_n & = {1\over 302400}(n+7)(n+6)(n+5)^2(n+4)^2(n+3)^2(n+2)(n+1) &\dimN\cr
& = 16,126,672,2772,9504,28314 \ldots
}$$

When $m=n-1$ is odd, the tensor ${\cal T}{}^{\a_1 \ldots \a_m}_{\b_1 \ldots \b_m}$ can be obtained from \Tndef\
by contracting a pair of indices; ${\cal T}{}^{\a_1 \ldots \a_m}_{\b_1 \ldots \b_m} ={\cal T}{}^{\a_1 \ldots \a_m \a_{m+1}}_{\b_1 \ldots \b_m \a_{m+1}}$.

The explicit expressions for the first few tensors read as follows,
\eqnn\trac
$$\eqalignno{
{\cal T}{}^{\a_1}_{\b_1} &= {1\over 16} \d^{\a_1}_{\b_1}\,,\cr
{\cal T}{}^{\a_1\a_2}_{\b_1\b_2} &=
{1\over 126}\Big[
\d^{(\a_1}_{\b_1}\d^{\a_2)}_{\b_2}
- {1\over 16}(\g\cdot\g)^{\a_1\a_2}_{\b_1\b_2}
\Big]\,,&\trac\cr
{\cal T}{}^{\a_1\ldots \a_3}_{\b_1 \ldots\b_3} &=
{1\over 672}\Big[
\d^{(\a_1}_{\b_1} \cdots \d^{\a_3)}_{\b_3}
- {3\over 20}\d^{(\a_1}_{(\b_1}(\g\cdot\g)^{\a_2\a_3)}_{\b_2\b_3)}
\Big]\,,\cr
{\cal T}{}^{\a_1\ldots \a_4}_{\b_1 \ldots\b_4} &=
{1\over 2772}\Big[ \d^{(\a_1}_{\b_1} \cdots \d^{\a_4)}_{\b_4}
- {1\over 4}\d^{(\a_1}_{(\b_1}\d^{\a_2}_{\b_2} (\g\cdot\g)^{\a_3\a_4)}_{\b_3\b_4)}
+ {1\over 160}(\g\cdot\g)^{(\a_1\a_2}_{(\b_1\b_2}(\g\cdot\g)^{\a_3\a_4)}_{\b_3\b_4)}
\Big]\,.\cr
}$$
Using the integration formula of \refs{\humberto,\coefftwo} and the above $\g$-traceless tensors
it follows that\foot{Note that all numbers in \dladlb\ have a geometrical meaning. The number 8 is the ghost anomaly
(the first Chern class of the projective pure spinor space), 11 is the complex dimension of the pure spinor space
and 12 is the degree of the projective pure spinor space \refs{\humberto,\character}.}
\eqn\dladlb{
\int[d\l][d\lb]e^{-(\l\lb)}(\l\lb)^m \l^{\a_1}\cdots \l^{\a_n}\lb_{\b_1}\cdots \lb_{\b_n} =
\({A_g\over 2\pi}\)^{\mkern-6mu 11}{12\,\Gamma(8+m+n)\over \Gamma(11)}{\cal T}{}^{\a_1 \ldots\a_n}_{\b_1 \ldots\b_n}.
}
To see this it is enough to check that the right-hand side of \dladlb\ has the same symmetries of the left-hand side
and it is correctly normalized.

Let us define the tensor ${\bar T}^{\a\b\g;\s_1 \ldots\s_5}$ by \multiloop
\eqn\defTf{
\lb_\a\lb_\b\lb_\g {\bar T}^{\a\b\g;\s_1 \ldots\s_5} = {\bar T}^{\s_1 \ldots\s_5},
}
where ${\bar T}^{\s_1 \ldots\s_5}$ is given in \Ttensors.
Since one can take ${\bar T}^{\a\b\g;\s_1 \ldots\s_5}$ to be $\g$-traceless in the $(\a\b\g)$ indices
it follows from \trac\ that
\eqn\fol{
{\cal T}{}^{\a_1\a_2\a_3}_{\b_1\b_2\b_3} {\bar T}^{\b_1\b_2\b_3;\s_1 \ldots\s_5}  = {1\over 672} {\bar T}^{\a_1\a_2\a_3;\s_1 \ldots\s_5}.
}
\proclaim Theorem 1. Let $f(\l^{n+3},\lb^n,\t)$ be a general superfield with ghost-number $+3$, then
\eqn\trick{
\bigl\langle f(\l^{n+3},\lb^n,\t)\bigr\rangle_{(m,g)} = \bigl\langle \llb^n {\hat f}(\l^3,\t^5)\bigr\rangle_{(m,g)}
}
where
$$
{\hat f}(\l^3,\t^5) = 672\, \l^{\b_1}\l^{\b_2}\l^{\b_3}\, {\cal T}{}_{\b_1 \ldots\b_{n+3}}^{\s_1 \ldots\s_{n+3}}\,
f_{\s_1 \ldots\s_{n+3};\d_1 \ldots\d_5}^{\b_4 \ldots\b_{n+3}}\t^{\d_1} \ldots\t^{\d_5}.
$$

\noindent{\sl Proof}.
Integrating the right-hand side of \trick\ over $[dr]$ and $[d\t]$ using the measures of \measures\ and the
definition \save\ yields
$$\eqalignno{
{\rm RHS} = 11!\,5!\,c_rc_\t \int [d\l][d\lb]& e^{-\llb}\, 672\llb^{n+m-3}
\l^{\b_1}\l^{\b_2}\l^{\b_3}\,\lb_{\g_1}\lb_{\g_2}\lb_{\g_3}\cr
&\times {\bar T}^{\g_1\g_2\g_3;\d_1 \ldots\d_5}
{\cal T}{}_{\b_1 \ldots\b_{n+3}}^{\s_1 \ldots\s_{n+3}}\,
f_{\s_1 \ldots\s_{n+3};\d_1 \ldots\d_5}^{\b_4 \ldots\b_{n+3}}.
}$$
Given that the ${\cal T}$ tensors are normalized such that ${\cal T}{}^{\a_1 \ldots\a_p}_{\b_1 \ldots\b_p} = 1$ the integration over
the pure spinors $\l$ and $\lb$ using \dladlb\ leads to
\eqnn\easyto
$$\eqalignno{
{\rm RHS} & = 11!\,5!\,c_rc_\t \({ A_g\over 2\pi}\)^{\!11}{\Gamma(8+m+n)\over 302400} \, 672
{\cal T}{}^{\b_1\b_2\b_3}_{\g_1 \g_2 \g_3}
{\bar T}^{\g_1\g_2\g_3;\d_1 \ldots\d_5}
{\cal T}{}_{\b_1 \ldots\b_{n+3}}^{\s_1 \ldots\s_{n+3}}\,
f_{\s_1 \ldots\s_{n+3};\d_1 \ldots\d_5}^{\b_4 \ldots\b_{n+3}}\cr
& = 11!\,5!\,c_rc_\t \({ A_g\over 2\pi}\)^{\!11}{\Gamma(8+m+n)\over 302400} \,
{\bar T}^{\b_1\b_2\b_3;\d_1 \ldots\d_5}
{\cal T}{}_{\b_1 \ldots\b_{n+3}}^{\s_1 \ldots\s_{n+3}}\,
f_{\s_1 \ldots\s_{n+3};\d_1 \ldots\d_5}^{\b_4 \ldots\b_{n+3}} &\easyto
}$$
where \fol\ has been used in the second line. However it is easy to show that the evaluation of the left-hand side
of \trick\ is equal to \easyto, finishing the proof.

For completeness, note that ${\bar T}^{\a\b\g;\s_1 \ldots\s_5}$ defined in \defTf\ is proportional to the pure
spinor correlator $\langle \l^\a\l^\b\l^\g \t^{\s_1} \ldots \t^{\s_5}\rangle_{(n,g)}$. Indeed, a short computation shows
that
\eqn\origDef{
\langle \l^\a\l^\b\l^\g \t^{\s_1} \ldots \t^{\s_5}\rangle_{(n,g)}
= \halfap2\Big({2\pi\over A_g}\Big)^{\!\! 5/2} {\Gamma(8+n)\over 302400}{R\over 672}\,{\bar T}^{\a\b\g;\s_1 \ldots\s_5}\,.
}
As a consistency check, multiplying both sides by $\g^m_{\a\d_1}\g^n_{\b\d_2}\g^p_{\g\d_3}(\g_{mnp})_{\d_4\d_5}$ recovers \Ndef,
$$
\langle (\l\g^m\t)(\l\g^n\t)(\l\g^p\t)(\t\g_{mnp}\t)\rangle_{(n,g)} =
\halfap{2} \Big({ 2\pi\over A_g}\Big)^{\mkern-6mu 5/2} 2^7 R\,{\Gamma(8+n)\over 7!}
$$
where we used that ${\bar T}^{\a\b\g;\d_1 \ldots\d_5} \g^m_{\a\d_1}\g^n_{\b\d_2}\g^p_{\g\d_3}(\g_{mnp})_{\d_4\d_5} = 5160960$ \stahnCorr.

As an example, the
function $f(\l^4,\lb,\t) \equiv \l^{\a_1}\l^{\a_2}\l^{\a_3}\l^{\a_4}\lb_{\b_1}
f^{\b_1}_{\a_1\a_2\a_3\a_4}(\t)$ can be easily rewritten according to the Theorem 1 by using
\eqnn\trickQQ
$$\eqalignno{
672\,\l^{\s_1}\l^{\s_2}\l^{\s_3} {\cal T}^{\a_1\a_2\a_3\a_4}_{\s_1\s_2\s_3\b_1} &= {2\over 33}\Big[
\big\{\d_{\b_1}^{\a_1}\la2\la3\la4 + (\a_1\leftrightarrow \a_2,\a_3,\a_4)\big\} &\trickQQ \cr
& - {1\over 12}(\l\g^m)_{\b_1}\big\{\g_m^{\a_1\a_2}\la3\la4 + \g_m^{\a_1\a_3}\la2\la4 + \g_m^{\a_1\a_4}\la2\la3 \cr
& \qquad\qquad\qquad+ \g_m^{\a_2\a_3}\la1\la4 + \g_m^{\a_2\a_4}\la1\la3 + \g_m^{\a_3\a_4}\la1\la2\big\}\Big]\,.
}$$

\medskip
\noindent{\it A.1 Factoring $\llb^6$ from  $L^m_{1234}(\l,\lb,r)$ and $T_{12,3,4}(\l,\lb,r)$}
\medskip

\noindent Because of the constraint \PSconstraints\ the
definition \Tijdef\ can be written as
\eqnn\factorL
$$\eqalignno{
L_{1234}^x(\l,\lb,r) &=  (\lb\g^a \g^b \g^c r) (\lb\g^{def} r) (\l\g^{adefm}\l)(\lb\g^n \g^m \g^p r) (\lb\g^{qrs} r) (\l\g^{nqrsu}\l)\cr
& \times (\lb\g^t \g^u \g^v r) (\lb\g^{ghi} r) (\l\g^{tghib}\l)\big[ (\l\g^c W_1)(\l\g^p W_2)(\l\g^v W_3) A_4^x\big] &\factorL
}$$
Applying the identity $(\lb\g_{mnp}r)(\l\g^{amnpb}\l) = 48(\l\lb)(\l\g^{a}\g^b r) - 48(\l\g^a\g^b\lb)(\l r)$
and using the pure spinor constraint gives
\eqnn\LagaD
$$\eqalignno{
L_{1234}^x(\l,\lb,r) &=  - 48^3 \llb^3 (\l\g^d\g^a r) (\l\g^g\g^e r) (\l\g^b\g^i r) &\LagaD\cr
&\quad\times (\lb\g^a \g^c \g^b r)(\l\g^c W_1) (\lb\g^e \g^f \g^d r) (\l\g^f W_2) (\lb\g^i\g^h\g^g r) (\l\g^h W_3)A^x_4\cr
}$$
where we also renamed indices.
Using
\eqn\repeat{
(\lb\g^{m}\g^{p}\g^{n}r)(\l\g^{p}W^2) =
- (\l\g^{p}\g^{n}r)(\lb\g^{m}\g^{p}W^2)
- (\l\g^{p}\g^{m}\lb)(r\g^{n}\g^{p}W^2),
}
in the last three factors and doing straightforward algebra yields,
\eqnn\bTh
$$\eqalignno{
L^x_{1234}(\l,\lb,r) &= 48^3\,8 \llb^5 Q\Bigl[ &\bTh \cr
&\quad + (\l\g^a\g^d r)(\l\g^c\g^e r)(\l\g^b\g^f r)(r\g^{ab}W^1)(r\g^{de}W^2)(\lb\g^{cf}W^3)\cr
&\quad + (\l\g^a\g^d r)(\l\g^c\g^e r)(\l\g^b\g^f r)(r\g^{ab}W^1)(\lb\g^{de}W^2)(r\g^{cf}W^3)\cr
&\quad + (\l\g^a\g^d r)(\l\g^c\g^e r)(\l\g^b\g^f r)(\lb\g^{ab}W^1)(r\g^{de}W^2)(r\g^{cf}W^3)\cr
&\quad - (\l\g^c\g^e r)(\l\g^b\g^f r)(\l\g^a\g^d \lb) (r\g^{ab}W^1)(r\g^{de}W^2)(r\g^{cf}W^3)\cr
&\quad + (\l\g^c\g^e r)(\l\g^a\g^d r)(\l\g^b\g^f \lb) (r\g^{ab}W^1)(r\g^{de}W^2)(r\g^{cf}W^3)\cr
&\quad + (\l\g^a\g^d r)(\l\g^b\g^f r)(\l\g^c\g^e \lb) (r\g^{ab}W^1)(r\g^{de}W^2)(r\g^{cf}W^3)\Bigr]\cr
& + 48^3\,8 \llb^6 (\l\g^a\g^d r)(\l\g^c\g^e r)(\l\g^b\g^f r) (r\g^{ab}W^1)(r\g^{de}W^2)(r\g^{cf}W^3)\cr
}$$
It is easy to check that \bTh\ is totally antisymmetric in $[123]$ as required.
The terms proportional to $\llb^5\,Q$ will be rewritten using \trickQQ\ and we identified $(\l r) = Q$ because of
the factor $e^{-(r\t)}$ in $\cal N$.
Despite the explicit appearance of the BRST charge in some terms, they are not
BRST-trivial because of the remaining factor $\lb$. However, since
$Q^2=0$ and the difference between $(\l\g^a\g^d r)$ and $(\l\g^{ad} r)$ is proportional to $Q$, one
can replace all factors of $(\l\g^a\g^d r)$ by $(\l\g^{ad} r)$. Doing this replacement
is also allowed in the last term because there are no factors of $\lb_\a$, so the BRST charge vanishes in the cohomology.
Similarly, $(\l\g^a\g^d\lb)$ can
be substituted by $(\l\g^{ad}\lb)$ since the difference is BRST-trivial due to the resulting factor of $Q\llb^6$.
Therefore \bTh\ becomes
\eqn\bThD{
L^x_{1234}(\l,\lb,r) = 48^3\,8 \bigl( F_{123} + F_{312} + F_{231} - G_{123} - G_{312} - G_{231} + H_{123}\bigr)A^x_4
}
where
\eqnn\Fdef
\eqnn\Gdef
\eqnn\Hdef
$$\eqalignno{
F_{123} &= \llb^5 Q(\l\g^{ad} r)(\l\g^{ce} r)(\l\g^{bf} r)(r\g^{ab}W^1)(r\g^{de}W^2)(\lb\g^{cf}W^3)\,,&\Fdef\cr
G_{123} &= \llb^5 Q(\l\g^{ce} r)(\l\g^{bf} r)(\l\g^{ad} \lb) (r\g^{ab}W^1)(r\g^{de}W^2)(r\g^{cf}W^3)\,,&\Gdef\cr
H_{123} &= \llb^6 (\l\g^{ad} r)(\l\g^{ce} r)(\l\g^{bf} r) (r\g^{ab}W^1)(r\g^{de}W^2)(r\g^{cf}W^3)\,.&\Hdef\cr
}$$
It is not difficult to show that $H_{123}$ is totally antisymmetric in $[123]$ whereas $F_{123}$ and $G_{123}$
are antisymmetric in $[12]$.

Let us rewrite the superfield $F_{123}$ using the Theorem~1. Since the $\g$-matrix traceless tensors
are normalized such that ${\cal T}^{\a_1 \ldots\a_n}_{\a_1 \ldots\a_n} = 1$, the factor $\llb^5$ is inert under the
application of the theorem and
one can use \trickQQ\ directly. Furthermore, all terms which still contain
an explicit BRST charge after using \trickQQ\ will be BRST-trivial because of the factor $\llb^6$. So in fact only
four terms in \trickQQ\ are non-vanishing when applied to $F_{123}$.
After straightforward algebra and discarding BRST-exact terms,
\eqnn\FtheoT
$$\eqalignno{
F_{123} = {2\over 33}&\llb^6\Bigl[ - (\l\g^{ad}r)(\l\g^{ce}r)(\l\g^{bf}r)(r\g^{ab}W^1)(r\g^{de}W^2)(r\g^{cf}W^3) \cr
& - {1\over 4}(\l\g^{de}r)(\l\g^{af}r)(\l\g^{bc}r)(r\g^{ab}W^3)(r\g^{cd}W^1)(r\g^{ef}W^2)\cr
& + {1\over 4}(\l\g^{de}r)(\l\g^{af}r)(\l\g^{bc}r)(r\g^{ab}W^3)(r\g^{cd}W^2)(r\g^{ef}W^1)\Bigr] &\FtheoT\cr
}$$
which implies that $F_{123} = - {1\over 11} H_{123}$ (and similarly
$G_{123} = {1\over 11} H_{123}$).
Plugging these results into \bThD\ and taking into account the total antisymmetry of $H_{ijk}$ one finally obtains
$L_{123}=(48^3\,40/11) H_{123}$. Identical manipulations apply to $T_{12,3,4}$, so
\eqn\Lfinal{
\eqalign{
L_{1234}^m(\l,\lb,r) &= {48^3\,40\over 11}\llb^6 (\l\g^{af} r)(\l\g^{bc} r)(\l\g^{de} r) (r\g^{ab}W^1)(r\g^{cd}W^2)(r\g^{ef}W^3) A_4^m\,,\cr
T_{12,3,4}(\l,\lb,r) &= {48^3\,40\over 11}\llb^6 (\l\g^{af} r)(\l\g^{bc} r)(\l\g^{de} r) (r\g^{ab}W_{12})(r\g^{cd}W_3)(r\g^{ef}W_4)\,.
}}
Similar manipulations can be used in $S^m_{1234}(\l,\lb,r)$ but for historic reasons we computed the five covariant derivatives
before rewriting it with the factor $\llb^6$. The resulting expression is not particularly illuminating and was therefore
omitted.

\appendix{B}{Integration by parts}
\applab\IBPappendix

\noindent
Noting that one can replace $\Delta(z_j;z_k;z_l)X_{1j}$ by $\Delta(z_1;z_k;z_l)X_{1j}$ in \calTdef\ because
$U_iU_j\sim (dW_{ij})(z_j)\eta_{ij} = (dW_{ij})(z_i)\eta_{ij}$
it is straightforward to show that the identities which eliminate $X_{1i}$ and $\bar X_{1j}$ are given by
\eqnn\prots
$$\eqalignno{
\Delta(1,3,4)\bar\Delta(2,3,4)X_{12}\bar X_{12}&= \Delta(1,3,4)\bar\Delta(1,3,4)(X_{23}+X_{24})(\bar X_{23} + \bar X_{24})\cr
&\quad{}+ \Delta(1,3,4)\bar\Delta(2,3,4) s_{12} \tilde\Omega_{21}
- \Delta(1,2,4)\bar\Delta(1,3,4)s_{23} \tilde\Omega_{32}\cr
&\quad{}+\Delta(1,2,3)\bar\Delta(1,3,4) s_{24} \tilde\Omega_{42}\,,\cr
\Delta(1,3,4)\bar\Delta(1,2,4)X_{12}\bar X_{13} &= (X_{23} + X_{24})(-\bar X_{23} + \bar X_{34})\Delta(1,3,4)\bar\Delta(1,2,4)\cr
&\quad{} + s_{23} \tilde\Omega_{23}\Delta(1,3,4)\bar\Delta(1,2,4)\,,\cr
\Delta(1,3,4)\bar\Delta(1,2,4)X_{12}\bar X_{23} &= \left[(X_{23} + X_{24}) \bar X_{23} - s_{23} \tilde\Omega_{23}\right]\Delta(1,3,4)\bar\Delta(1,2,4)\,,\cr
\Delta(1,3,4)\bar \Delta(1,2,3)X_{12}\bar X_{34} &= (X_{23} + X_{24}) \bar X_{34}\Delta(1,3,4)\bar \Delta(1,2,3), &\prots\cr
}$$
where we used that
\eqn\dOmega{
\pb_i X_{ji} =  s_{ij}\tilde\Omega_{ji}, \quad \p_i\bar X_{ji} =  s_{ij}\tilde\Omega_{ij}, \quad \pb_i X_{ij} = - s_{ij}\tilde\Omega_{ji},
\quad \p_i\bar X_{ij} = - s_{ij}\tilde\Omega_{ij}\,,
}
and defined $\tilde\Omega_{ij}=(\ap/2) \pi\Omega(z_i,z_j)$. All other identities needed to write $|{\cal F}|^2$ in a basis
of integrals follow from the above by relabeling.
Applying them together with
\eqn\iomegaij{
\tilde\Omega_{ij}\Delta(j,k,l)\bar\Delta(i,k,l)=
\tilde\Omega_{ji}\Delta(i,k,l)\bar\Delta(j,k,l)\,,
}
implies that $|{\cal F}|^2$ is equal to
\eqnn\UbbCs
$$\eqalignno{
& + |C_{32,1,4}|^2 (X_{23}\bar X_{23} - s_{23}\tilde\Omega_{23})\Delta(1,3,4)\bar\Delta(1,2,4) &\UbbCs\cr
& - |C_{24,1,3}|^2 (X_{24}\bar X_{24}- s_{24}\tilde\Omega_{24})\Delta(1,3,4)\bar\Delta(1,2,3)\cr
& + |C_{34,1,2}|^2 (X_{34}\bar X_{34} - s_{34}\tilde\Omega_{34})\Delta(1,2,4)\bar\Delta(1,2,3)\cr
&+ C_{32,1,4}\tilde C_{24,1,3} X_{23}\bar X_{24}\Delta(1,2,4)\bar\Delta(1,3,4) + C_{32,1,4}\tilde C_{34,1,2} X_{23}\bar X_{34}\Delta(1,3,4)\bar\Delta(1,2,4)\cr
&+ C_{24,1,3}\tilde C_{32,1,4} X_{24}\bar X_{23}\Delta(1,3,4)\bar\Delta(1,2,4) + C_{24,1,3}\tilde C_{34,1,2} X_{24}\bar X_{34}\Delta(1,3,4)\bar\Delta(1,2,3)\cr
&+ C_{34,1,2}\tilde C_{32,1,4} X_{34}\bar X_{23}\Delta(1,2,4)\bar\Delta(1,3,4) + C_{34,1,2}\tilde C_{24,1,3} X_{34}\bar X_{24}\Delta(1,2,4)\bar\Delta(1,3,4)\cr
&+ s_{12}\tilde\Omega_{12} |M_{12,3,4}|^2 \Delta(2,3,4)\bar\Delta(1,3,4) - s_{13}\tilde\Omega_{13} |M_{13,2,4}|^2 \Delta(2,3,4)\bar\Delta(1,2,4)\cr
&+ s_{14}\tilde\Omega_{14} |M_{14,2,3}|^2 \Delta(2,3,4)\bar\Delta(1,2,3) + s_{23}\tilde\Omega_{23} |M_{23,1,4}|^2 \Delta(1,3,4)\bar\Delta(1,2,4)\cr
&- s_{24}\tilde\Omega_{24} |M_{24,1,3}|^2 \Delta(1,3,4)\bar\Delta(1,2,3)
 + s_{34}\tilde\Omega_{34} |M_{34,1,2}|^2 \Delta(1,2,4)\bar\Delta(1,2,3)
}$$
where we defined (the others follow from relabeling)
\eqn\Cthreefour{
C_{24,1,3} \equiv  M_{24,1,3} + M_{14,2,3} + M_{12,3,4}.
}

The $\ap$-expansion of the above integrals has not been derived but one can argue from the results of \FiveSduality\ that
$\eta_{ij}\bar\eta_{ij} - \tilde\Omega_{ij}s_{ij}^{-1}$ and $\eta_{ij}\bar\eta_{ik}$ have no kinematic poles. Therefore the
leading-order contribution from \UbbCs\ is given by the $\tilde\Omega_{ij}$ terms and it follows from
relabeling of integration variables that they are all equal to $\pm\int\tilde\Omega_{12}\Delta(2,3,4)\bar\Delta(1,3,4)$ (the
sign is easy to obtain). Thus the low-energy limit of $|{\cal F}|^2$ in \Athree\ corresponds to\foot{The minus sign compensates
the ``convention'' $ik^m \rightarrow k^m$ in the Koba--Nielsen factor.}
\eqn\Abbprime{
-{\sqrt{2}\pi^7\over 2^{23}\,3^7} \kappa^4 e^{4\l}\halfap7
\int {d^2\Omega_{IJ} \over Z_3^{-10}}\int\prod_{i=1}^4 d^2z_i\, \Omega_{12}\Delta(2,3,4)\bar\Delta(1,3,4)
\bigl\langle {\cal K} \bigr\rangle_{(-3)}\eikx4
}
where we used $\tilde\Omega_{12} = \pi(\ap/2)\Omega_{12}$ and defined,
\eqn\KfinalB{
{\cal K} =   {|T_{23,1,4}|^2\over s_{23}} + {|T_{24,1,3}|^2\over s_{24}} + {|T_{34,1,2}|^2\over s_{34}}
   + {|T_{12,3,4}|^2\over s_{12}} + {|T_{13,2,4}|^2\over s_{13}} + {|T_{14,2,3}|^2\over s_{14}}\,.
}

\medskip
\noindent{\it B.1 Open superstring}
\medskip

\noindent
In the case of the open superstring it is not difficult to argue that the corresponding low-energy limit is
\eqn\Kopen{
{\cal K}^{(\rm open)} 
={T_{23,4,1} + T_{41,2,3} \over s_{23}} + {T_{34,1,2} + T_{12,3,4}\over s_{34}}.
}
The component expansion of \Kopen\ provides a good consistency check for the methods of \PSS\ since
one recovers the $\ap^3$ interaction of the open superstring tree-level amplitude \treebbI,
\eqn\OK{
{\langle T_{23,4,1} + T_{41,2,3}\rangle \over s_{23}} + {\langle T_{34,1,2} + T_{12,3,4}\rangle \over s_{34}} =
1344\cdot 40\cdot48^3\cdot 2880\,A^{\rm YM}_{1234} s_{12}s_{13}s_{23}.
}

\listrefs

\bye